\documentstyle[aps,amssymb,eqsecnum,preprint,tighten,psfig,epsf]{revtex}
\begin{document}
\draft
\title{\rightline{\small{\em To appear in Phys. Rev. D, July 15 (2001)\/}}
       Complete null data for a black hole collision }

\author{Roberto G\'omez}
\address{Department of Physics and Astronomy,
         University of Pittsburgh, Pittsburgh, Pennsylvania 15260}
\author{Sascha Husa and Jeffrey Winicour}
\address{Department of Physics and Astronomy,
         University of Pittsburgh, Pittsburgh, Pennsylvania 15260 \\
     and Max-Planck-Institut f\" ur Gravitationsphysik,
         Albert-Einstein-Institut, 14476 Golm, Germany}
\maketitle

\begin{abstract}

We present an algorithm for calculating the complete data on an event horizon
which constitute the  necessary input for characteristic evolution of the
exterior spacetime. We apply this algorithm to study the intrinsic and extrinsic
geometry of a binary black hole event horizon, constructing a sequence of
binary black hole event horizons which approaches a single Schwarzschild black
hole horizon as a limiting case. The linear perturbation of the Schwarzschild
horizon provides global insight into the close limit for binary black holes, in
which the individual holes have joined in the infinite past. In general there
is a division of the horizon into interior and exterior regions, analogous to
the division of the Schwarzschild horizon by the $r=2M$ bifurcation sphere. In
passing from the perturbative to the strongly nonlinear regime there is a
transition in which the individual black holes persist in the exterior portion
of the horizon. The algorithm is intended to provide the data sets for
production of a catalog of nonlinear post-merger wave forms using the PITT null
code.

\end{abstract}


\section{Introduction}

In previous work, we developed a model which generates the intrinsic null
geometry of an event horizon with the ``pair-of-pants'' structure
characteristic of a binary black hole merger~\cite{ndata,asym}. In this paper,
we extend this approach to determine all extrinsic curvature properties of such
horizons, thus providing a complete stand-alone description of the event
horizon of binary black holes. We apply this work to study the event horizon of
a head-on collision of black holes using a sequence of models which embraces 
not only the perturbative regime of the close approximation~\cite{pp}, where
the merger takes place in the distant past, but also includes the highly
nonlinear regime. In the perturbative regime the individual black holes merge
in an interior region of the horizon, corresponding to the region of the
Schwarzschild event horizon lying inside the $r=2M$ bifurcation sphere. But we
show how dramatically nonlinear effects can push the merger into the exterior
portion of the horizon. 

Beyond the investigation of the horizon geometry of a black hole collision, the
major motivation for this work is to provide the null data necessary to compute
the emitted gravitational wave by means of a characteristic evolution of the
exterior spacetime. In the Cauchy problem, the necessary data on a spacelike
hypersurface are the intrinsic metric and extrinsic curvature, subject to
constraints. On a null hypersurface, such as an event horizon, the situation is
quite different. The necessary null data consist of the conformal part of the
intrinsic (degenerate) metric, which can be given freely as a function of
the affine parameter. Then the surface area of the horizon is determined, via an
ordinary differential equation along the null rays (the Raychaudhuri equation),
in terms of an integration constant supplied by the mass of the final black
hole. Similarly, all extrinsic curvature components of the horizon are
determined by ordinary differential equations in terms of integration constants
supplied by the final black hole. Whereas in principle the intrinsic conformal
geometry is the only data on the horizon necessary for the characteristic
initial value problem, in practice the surface area and extrinsic curvature are
essential to supply the start-up data for the implementation of a
characteristic evolution code, such as the PITT null code~\cite{high,wobb}.
The main results of this paper concern the understanding of the nonlinear
nature of the underlying ordinary differential equations from geometrical,
physical and numerical points of view. 

\begin{figure}
\centerline{\epsfxsize=6.0in\leavevmode\epsfbox{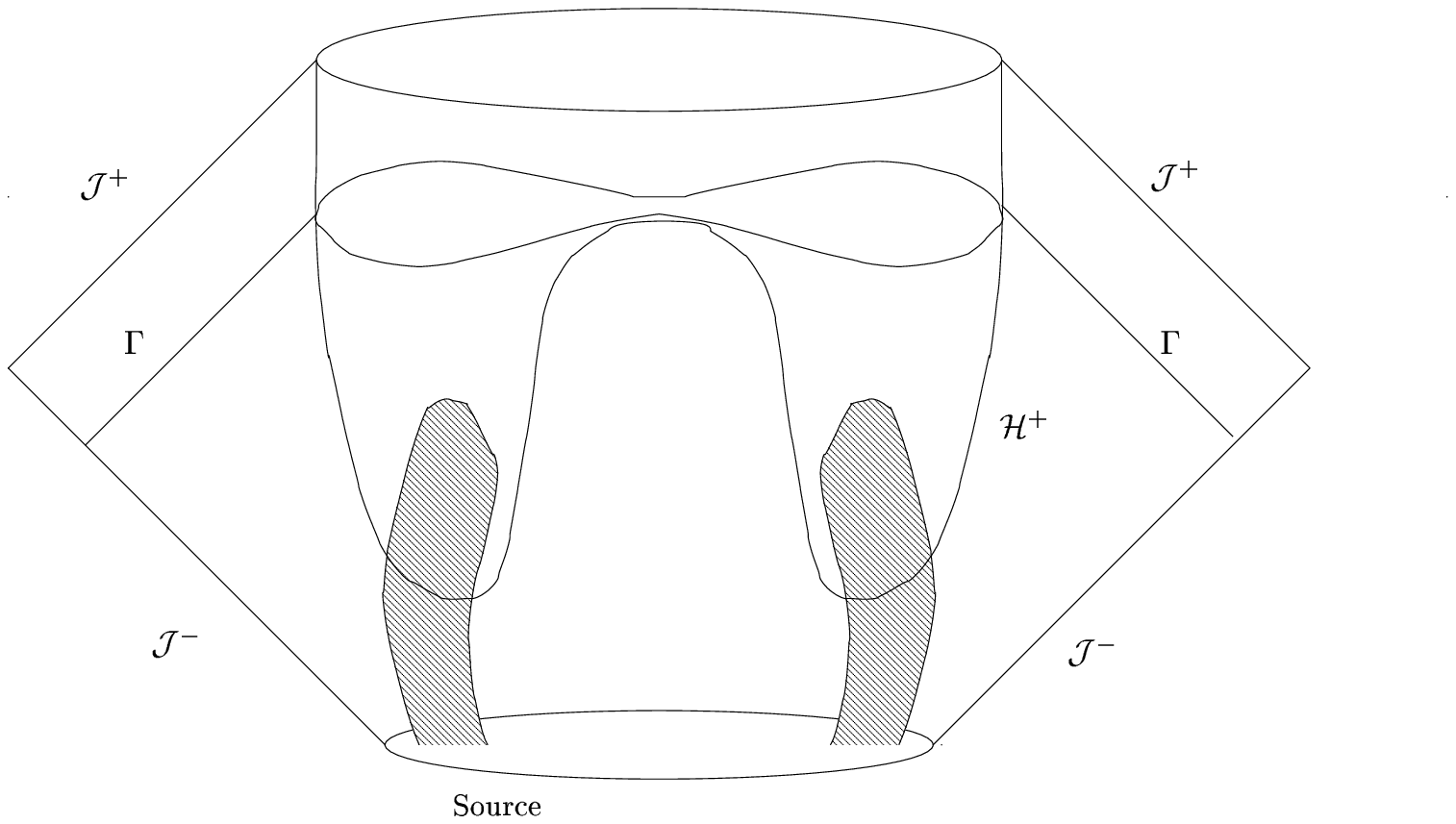}}
\caption{The global binary black hole problem involves the event horizon
${\cal H}^+$ and future and past null infinity, here approximated
by ${\cal J}^+$ and ${\cal J}^-$. The sources which formed the black holes,
e.g. collapsing matter or imploding gravitational waves, are indicated
by the shaded regions. The sources are surrounded by a null world tube $\Gamma$,
outside of which characteristic evolution supplies the exterior spacetime. }
\label{fig:strategy}
\end{figure}

Given the intrinsic geometry and extrinsic curvature of the horizon, the
strategy behind the characteristic approach to the computation of the emitted
wave has been outlined elsewhere~\cite{kyoto}. It consists of two evolution
stages based upon the double null initial value problem. Referring to Fig.
\ref{fig:strategy}, the necessity of two stages results from the disconnected
nature of the two null hypersurfaces on which boundary conditions must be
satisfied: (i) the event horizon ${\cal H}^+$, where binary black hole data is
prescribed, and (ii) past null infinity ${\cal I}^-$, where ingoing radiation
must be absent, at least in the late stage to the future of the hypersurface
$\Gamma$ which is decoupled from the formation of the individual black holes.
Rather than directly attempting to solve this mixed version of a characteristic
initial value problem, the stage I evolution is based upon two intersecting
null hypersurfaces consisting of the event horizon, where binary black hole
data is prescribed, and an ingoing null hypersurface ${\cal J}^+$ approximating
future null infinity ${\cal I}^+$, where data corresponding to no {\em
outgoing} radiation is prescribed. The characteristic evolution then proceeds
{\em backward in time} along ingoing null hypersurfaces extending to ${\cal
I}^-$ to determine the spacetime exterior to an ingoing null hypersurface
$\Gamma$. As indicated in the figure, the evolution terminates at $\Gamma$
because the horizon splits into two individual black holes. This stage I
solution is the {\em advanced} solution to the problem in the sense that
radiation from ${\cal I}^-$ is absorbed by the black holes but no outgoing wave
is radiated to ${\cal I}^+$. Stage II provides the {\it retarded} solution,
where outgoing but no ingoing radiation is present outside $\Gamma$, by running
forward in time a double null evolution based upon the intersecting null
hypersurfaces $\Gamma$, where the stage I data is prescribed, and an outgoing
null hypersurface ${\cal J}^-$ approximating ${\cal I}^-$, where data
corresponding to no ingoing radiation is prescribed to the future of $\Gamma$.
The stage II evolution produces the retarded solution for the spacetime outside
the world tube $\Gamma$. The approach is a nonlinear version of the standard
method of determining the retarded solution $\Phi_{RET}$ of the linear wave
equation $\Box \Phi =S$, with source $S$, by first finding the advanced
solution $\Phi_{ADV}$ and then superimposing the source-free solution
$\Phi_{RET} - \Phi_{ADV}$. In the nonlinear case, where standard Green
function techniques cannot be used to define retarded and advanced solutions,
they can be defined by requiring the absence of radiation at ${\cal I}^+$
(retarded) or ${\cal I}^-$ (advanced). In the present case, the data on the
world tube $\Gamma$ represents the interior source which has led to the
formation of the individual black holes. (In a normal physical context, the
source consists of two stars undergoing gravitational collapse but in a purely
vacuum scenario imploding gravitational waves can play the role of the matter.)
The justification of this two stage approach is that it reduces to the standard
linear method in the close approximation where the geometry can be regarded as
a perturbation of a Schwarzschild background. A purely perturbative
characteristic treatment of the close approximation using the characteristic
approach has been carried out in a separate study and the advanced solution has
been successfully computed~\cite{close1}.

We intend to carry out the ``backward in time'' evolution using the PITT null
code~\cite{high,wobb}, which is based upon the Bondi-Sachs version of the
characteristic initial value problem~\cite{bondi,sachs}. The present code is
designed to evolve forward in time along a foliation of spacetime by outgoing
null hypersurfaces. In this paper, in order to apply it to the backward in time evolution of a
black hole horizon, we describe the merger of two black holes in
the time reversed scenario of a white hole horizon, in which the black hole
merger is now represented by the fission of a white hole. The post-merger era
of the black hole horizon then corresponds to the pre-fission era of the white
hole; and the proposed backward in time evolution of the black hole horizon to
determine the exterior spacetime corresponds to a forward in time evolution of
the white hole horizon. Our algorithm provides the complete
white hole data necessary to carry out this evolution.

The specific application in this paper is to the axisymmetric head-on collision
of two equal mass black holes. However, the algorithm is capable of generating
intrinsic geometry and extrinsic curvature of an arbitrary event horizon,
including the case of inspiraling binary black holes of non-equal mass. In
previous studies of the intrinsic geometry of non-axisymmetric horizons, the
approach has revealed new features of the generic collision of two black holes,
such as an intermediate toroidal phase which precedes the merger~\cite{asym}.
Here we apply the algorithm to gain new insight into the global behavior of the
extrinsic curvature properties of the head-on collision. 

The material in the paper divides into two types: (i) general features of a
binary horizon and (ii) the calculation of the null data on the horizon
necessary to initiate the PITT code. Material of the first type, which
describes how a binary horizon deviates from the close approximation in the
nonlinear regime, appears in Secs.~\ref{sec:horstruct} and~\ref{sec:confmod} -
\ref{sec:results} and can be read independently. This material is described
most conveniently in terms of Sachs coordinates, introduced in Sec.
\ref{sec:horstruct}. Material of the second type involves the transformation
from Sachs to Bondi-Sachs coordinates, introduced in Sec.~\ref{sec:Bondisachs}.
This material supplies all data necessary to compute, using the PITT code, the
fully nonlinear merger to ringdown wave form, which will be the subject of
future work. The details of implementing the horizon data algorithm as a finite
difference code are given in the Appendixes. An independent code designed for
an axisymmetric horizon is also described and has been used as an independent
test of the full code.

We retain the conventions of our previous papers~\cite{ndata,asym,high,wobb},
with only minor changes in notation where noted in the text. For brevity, we
frequently use the notation $f_{,x}=\partial_x f$ to denote partial derivatives
and $\dot f =\partial_u f$ to denote retarded time derivatives.

\section{Characteristic data on a horizon}

\subsection{Horizon structure}
\label{sec:horstruct}

\begin{figure}
\centerline{\epsfxsize=6.0in\leavevmode\epsfbox{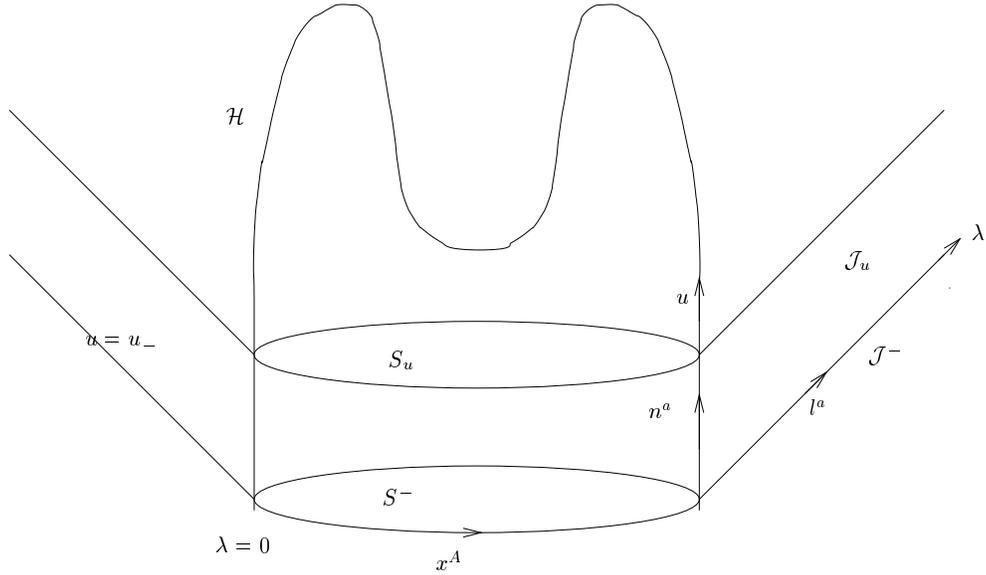}}
\caption{The white hole horizon ${\cal H}$ is foliated by an affine
parameter $u$, with ${\cal S}^-$ representing an early
quasi-stationary slice. The affine parameter $\lambda$
along the outgoing null hypersurfaces ${\cal J}_u$
emanating from the foliation is chosen so that $\lambda=0$ on ${\cal H}$.
The angular coordinates $x^A$ are chosen to be constant along the light
rays generating ${\cal H}$ and ${\cal J}_u$. The null vectors $l^a$ and $n^a$
are used to project tensor fields into ${\cal H}$.}
\label{fig:whole}
\end{figure}

The evolution of the exterior spacetime by the PITT code proceeds along a
family of outgoing null hypersurfaces. The characteristic initial value problem
for the evolution requires an inner boundary condition which can be set either
on a timelike world tube or, as a limiting case, on a null world tube. Here we
choose the inner boundary to be the null world tube representing a white hole
horizon ${\cal H}$. The white hole horizon pinches off in the future where its
generators either caustic or cross each other (such as at the vertex of a null
cone). As illustrated in Fig.~\ref{fig:whole}, we introduce (i) an affine null
coordinate $u$ along the generators of ${\cal H}$, which foliates ${\cal H}$
into cross sections ${\cal S}_u$ and labels the corresponding outgoing null
hypersurfaces ${\cal J}_u$ emanating from the foliation; (ii) angular
coordinates $x^A$ which are constant both along the generators of ${\cal H}$
and along the outgoing rays and (iii) an affine parameter $\lambda$ along the
outgoing rays normalized by $\nabla^{\alpha}u \nabla_{\alpha}\lambda =-1$, with
$\lambda =0$ on ${\cal H}$. In the resulting $x^{\alpha}=(u,\lambda ,x^A)$
coordinates, the metric takes the form
\begin{equation}
   ds^2  = -(W -g_{AB}W^A W^B)du^2
       -2dud\lambda -2g_{AB}W^Bdudx^A +  g_{AB}dx^Adx^B.
\label{eq:amet}
\end{equation}
The contravariant components are given by $g^{\lambda u}=-1$, $g^{\lambda
A}=-W^A$,  $g^{\lambda\lambda}=W$ and $g^{AB}g_{BC}=\delta^A_C$. In addition,
we set $g_{AB}=r^2h_{AB}$, where $\det(h_{AB})=\det(q_{AB})=q(x^A)$, where
$q_{AB}$ is some standard choice of unit sphere metric.

These coordinates were first introduced by Sachs to formulate the double-null
characteristic initial value problem~\cite{sachsdn}. The Bondi-Sachs coordinate
system ~\cite{bondi,sachs} used in the PITT code differs by the use of a
surface area coordinate $r$ along the outgoing cones rather than the affine
parameter $\lambda$. However, because the horizon is not a surface of constant
$r$, except in the special case of an ``isolated horizon''~\cite{ih5083}, it is
advantageous to first determine the necessary data in terms of an affine
parameter and later transform to the $r$-coordinate.

The requirement that ${\cal H}$ be null implies that $W=0$ on ${\cal H}$. There
is gauge freedom on ${\cal H}$ that we fix by choosing the shift so that
$\partial_u$ is tangent to the generators, implying that $W^A=0$ on ${\cal H}$;
and by choosing the lapse so that $u$ is an affine parameter, implying that
$\partial_\lambda W=0$ on ${\cal H}$. We adopt these choices throughout the
paper, and our results generally hold only on $\cal H$ ($\lambda = 0$) and when
these conditions are satisfied. Later, in Sec.~\ref{sec:headon}, we also fix
the
affine freedom in $u$ by specifying it on an initial slice ${\cal S}^-$ of
${\cal H}$, which is located at an early time approximating the asymptotic
equilibrium of the white hole at past time infinity $I^-$. The outgoing null
hypersurface ${\cal J}^-$ emanating from ${\cal S}^-$ approximates past null
infinity ${\cal I}^-$. In Sec. {}~\ref{sec:sminus}, we discuss the nature of
that approximation.

On ${\cal H}$, the affine tangent to the generators $n^a\partial_a=\partial_u$
(see Fig.~\ref{fig:whole}) satisfies the geodesic equation $n^b\nabla_b n^a=0$
and the hypersurface orthogonality condition  $n^{[a}\nabla^b n^{c]}=0$.
Following the approach of Refs.~\cite{ndata,asym}, we project 4-dimensional
tensor fields into ${\cal H}$ using the operator
\begin{equation}
       P_a^b = \delta_a^b + n_a l^b,
\end{equation}
where $l_a = -\nabla_a u$, and we use the shorthand notation $\perp T_a^b$ for
the projection (to the tangent space of ${\cal H}$) of the tensor field
$T_a^b$. The projection $\perp$ has gauge freedom corresponding to the choice
of affine parameter $u$. However, the action of $\perp$ on covariant indices is
independent of this freedom and equals the action of the pullback operator to
${\cal H}$.

In addition to the intrinsic geometry of ${\cal H}$, the necessary
characteristic data consist of the extrinsic curvature quantities given by
$\perp \nabla_a l_b$ (with gauge freedom corresponding to the affine choice
of $u$). Since $\nabla_a l_b -\nabla_b l_a =0$, the independent components are
determined by $\perp \nabla_{(a} l_{b)}$, which has the decomposition
\begin{equation}
  \perp \nabla_{(a} l_{b)} = -l_a \omega_b-\omega_a l_b +\tau_{ab}.
\label{eq:gradl}
\end{equation}
Here $\tau_{ab}$ describes the shear and expansion of the outgoing rays and
satisfies $n^a\tau_{ab}=0$ on ${\cal H}$. Following Hayward~\cite{haywsn}, we
call $\omega_a =\perp n^b \nabla_b l_a$ the twist of the affine foliation of
${\cal H}$. The twist also satisfies $n^a \omega_a =0$. Unlike $\tau_{ab}$, the
twist is an invariantly defined extrinsic curvature property of the $u=const$
cross sections of ${\cal H}$, independent of the boost freedom in the
extensions of $n_a$ and $l_a$ subject to the normalization $n^a l_a=-1$.

Note that it is natural geometrically to associate extrinsic curvature
properties of ${\cal H}$ with $(\perp \nabla_a)n^b$, in analogy with the
Weingarten map for a non-degenerate hypersurface~\cite{ih5083}. The
normalization $n^a l_a =-1$ then leads, via Eq.~(\ref{eq:gradl}), to
\begin{equation}
   l_b(\perp \nabla_a )n^b =-\omega_a.
\end{equation}
Thus the twist describes an extrinsic curvature property associated with $n^a$
as well as $l_a$. The other components of $(\perp \nabla_a )n^b$ are determined
by the shear and expansion of ${\cal H}$ (which vanish in the special case of
an isolated horizon).

We consider the double null initial value problem based upon data on the
horizon ${\cal H}$ and the outgoing null hypersurface ${\cal J}^-$ emanating
from ${\cal S}^-$, with the evolution proceeding along the outgoing null
hypersurfaces ${\cal J}_u$ emanating from the $u=const$ foliation ${\cal S}_u$.
In this problem, the complete (and unconstrained) characteristic data on ${\cal
H}$ are its affine parametrization $u$ and the conformal part $h_{AB}$ of its
degenerate intrinsic metric. Similarly, the characteristic data on ${\cal J}^-$
are its affine parametrization $\lambda$ and its intrinsic conformal metric
$h_{AB}$. The remaining data consist of the intrinsic metric and extrinsic
curvature of ${\cal S}^-$ (subject to consistency with the characteristic
data)~\cite{sachsdn,haywdn}. In Sachs coordinates, this data on ${\cal S}^-$
consists of $r$, $\dot r$, $\omega_a$ and $r_{,\lambda}$ (which determines the
expansion of the outgoing null rays) on ${\cal S}^-$. That completes the data
necessary to evolve the exterior spacetime. In carrying out such an evolution
computationally, the first step is to propagate the data given on ${\cal S}^-$
along the generators of ${\cal H}$ so that it can be supplied as boundary data
to the exterior evolution code. This first step is accomplished by means of
certain components of Einstein's equations.

Einstein's equation decompose into (i) hypersurface equations intrinsic to the
null hypersurfaces ${\cal J}_u$, which determine auxiliary metric quantities in
terms of the conformal metric $h_{AB}$; (ii) evolution equations determining
the rate of change $\partial_u h_{AB}$ of the conformal metric of ${\cal J}_u$;
and (iii) propagation equations which are constraints that need only be
satisfied on ${\cal H}$. One of the propagation equations is the ingoing
Raychaudhuri equation  $R_{uu}=0$ which determines the surface area variable
$r$ along the generators of ${\cal H}$ in terms of initial conditions on ${\cal
S}^-$ according to
\begin{equation}
      \ddot r =\frac{r}{8} \dot h^{AB} \dot h_{AB}.
\label{eq:ruu}
\end{equation}
The value of $\dot r$ on ${\cal S}^-$ measures the convergence of its ingoing
null rays and Eq.~(\ref{eq:ruu}) implies $\ddot r\le 0$. The remaining
propagation equations $R_{Au} =0$ propagate the twist $\omega_a$ along the
generators of ${\cal H}$.  Our coordinate conditions imply that
$\omega_u=\omega_{\lambda}=0$ and
\begin{equation}
        \omega_A=-\frac{1}{2}\partial_{\lambda}(g_{AB} W^B ). \label{eq:wA}
\end{equation}
The non-vanishing components propagate according to
\begin{equation}
 (r^2\omega_A)\dot {}=r^2 D_A(\frac{\dot r}{r})
            -\frac{1}{2}h^{BC}D_B(r^2 \dot h_{AC} ),
\label{eq:omegadot}
\end{equation}
where $D_A$ is the covariant derivative associated with $h_{AB}$. Here
$h^{AB}h_{BC}=\delta^A_C$ and $h^{AB}\dot h_{AB} =0$. Having determined $r$,
this equation can easily be integrated to determine $\omega_A$ on ${\cal H}$ in
terms of initial conditions on ${\cal S}^-$.  Once the propagation equations
are solved on ${\cal H}$ to determine $r$ and $\omega_a$, the Bianchi
identities ensure that they will be satisfied in the exterior spacetime as a
result of the ${\cal J}_u$ hypersurface equations and the evolution equations.

The propagation of $\tau_{ab}$ (the outward shear expansion and shear) along
${\cal H}$ requires the $R_{AB}=0$
components of Einstein equations, given by
\begin{eqnarray}
 R_{AB} &=& h_{AB} \bigg  (\frac{ {\cal R}}{2}-D^AD_A \,\log r\bigg )
          - D_A \omega_B - D_B \omega_A -2\omega_A\omega_B
               \nonumber \\
         &+& \frac{2}{r}(\omega_A \partial_B r +\omega_B \partial_A r
                          -h_{AB}h^{CD}\omega_C \partial_D r)
                 \nonumber \\
        &+& r^2\partial_\lambda \dot h_{AB}
              -\frac{r^2}{2} h^{CD}(\dot h_{AC}\partial_\lambda h_{BD}
                    +\dot h_{BD}\partial_\lambda h_{AC} )
                    \nonumber \\
         &+& 2(\dot r\partial_\lambda r +r\partial_\lambda \dot r)h_{AB}
        +r(\partial_\lambda r) \dot h_{AB} + r\dot r \partial_\lambda h_{AB}.
\label{eq:rab}
\end{eqnarray}
where ${\cal R}$ represents the curvature scalar of the metric $h_{AB}$. On
${\cal H}$, the equation $R_{AB}=0$ decomposes into the trace
\begin{equation}
\partial_u\partial_\lambda (r^2) = +D^AD_A \,\log r
-\frac{1}{2}{\cal R} +D^A\omega_A +h^{AB}\omega_A\omega_B
\label{eq:rlam}
\end{equation}
and, by introducing the dyad decomposition $h_{AB}=m_{(A}\bar
m_{B)}$, the trace-free part
\begin{equation}
     m^A m^B \bigg (r\partial_\lambda \partial_u(r h_{AB})-2D_A\omega_B
              -2\omega_A\omega_B +\frac{4}{r}\omega_A D_B r \bigg ) =0.
\label{eq:jlamdot}
\end{equation}
The trace equation propagates the outgoing expansion of the foliation ${\cal
S}_u$, as determined by $r_{,\lambda}$. The trace-free part is the evolution
equation for the data $h_{AB}$ on the foliation ${\cal J}_u$, applied at ${\cal
H}$ to evolve $h_{AB,\lambda}$, which describes the shear of the outgoing rays.
The outward shear constitutes part of the extrinsic curvature of ${\cal S}^-$.
Although its value is implicitly determined by the null data $h_{AB}$ on ${\cal
J}^-$, we view it as part of the initial data that must be specified on ${\cal
S}^-$.

In summary, the data for the double null problem includes the conformal metric
on ${\cal H}$ and the quantities $r$, $\dot r$, $\omega_a$, $r_{,\lambda}$ and
$h_{AB,\lambda}$ on ${\cal S}^-$. Equations~(\ref{eq:ruu}),
(\ref{eq:omegadot}), (\ref{eq:rlam}) and (\ref{eq:jlamdot}) are then used to
propagate the data on ${\cal S}^-$ to all of ${\cal H}$. The choice of
foliation of the horizon is an important but complicated aspect of this
problem. As in the case of Cauchy evolution, gauge freedom in the foliation
introduces arbitrariness into the dynamical description of any black hole
process and, in particular, the pair-of-pants structure underlying a black hole
merger. In the case of the horizon, the natural choice of an affine foliation
$u$ removes any time dependence from this gauge freedom but there remains the
affine freedom $u\rightarrow Au+B$, where $A$ and $B$ are ray dependent. In
Sec.~\ref{sec:headon}, we use the asymptotic equilibrium of the white hole as
$u\rightarrow -\infty$ to fix this freedom. We review in Sec.~\ref{sec:confmod}
how the data $h_{AB}$ and $r$ on ${\cal H}$ describing a white hole fission
(binary black hole merger) are provided by a conformal horizon
model\cite{ndata,asym}. The additional data required on ${\cal S}^-$ can be
inferred from the asymptotic properties of the white hole equilibrium at $I^-$,
as discussed in Sec.~\ref{sec:headon}. The evolution of the exterior spacetime
by the PITT code requires a transformation of the data on ${\cal H}$ to a
Bondi-Sachs coordinate system, as described next in Secs.~\ref{sec:Bondisachs}
-~\ref{sec:bondioff}.

\subsection{Bondi-Sachs coordinates}
\label{sec:Bondisachs}

The transformation to the Bondi-Sachs coordinates $x^{\alpha}=(u,r,x^A)$ used
in the PITT code consists in substituting the surface area coordinate $r$ for
the affine parameter $\lambda$. Since the horizon ${\cal H}$ does not in
general have constant $r$ it does not lie precisely on radial grid points.
Consequently, the assignment of horizon boundary values must be done on the
grid points nearest to the boundary. Thus an accurate prescription of boundary
conditions in the $r$-grid requires a Taylor expansion of the horizon data.

In Bondi-Sachs variables, the resulting metric takes the form
\begin{equation}
   ds^2=-\left(e^{2\beta}{V \over r} -r^2h_{AB}U^AU^B\right)du^2
        -2e^{2\beta}dudr -2r^2 h_{AB}U^Bdudx^A +r^2h_{AB}dx^Adx^B.
   \label{eq:umet}
\end{equation}
In relating this to the Sachs metric Eq.~(\ref{eq:amet}), it is
simplest to consider the contravariant form in which only the $g^{r\alpha}$
components differ. In terms of the corresponding metric variables,
\begin{eqnarray}
 g^{rr} &= e^{-2\beta} \frac{V}{r} &= (r_{,\lambda})^2 W
      - 2\, r_{,\lambda} r_{,A} W^{A} - 2\, r_{,\lambda} r_{,u}
   + \frac{r_{,A} r_{,B}}{r^2} h^{AB}
\label{eq:VH} \\
 g^{rA} &= -e^{-2\beta} U^A &=- r_{,\lambda} W^A + \frac{r_{,B}}{r^2} h^{AB}\\
 g^{ru} &=  -e^{-2\beta} &= - r_{,\lambda} .
\label{eq:rl}
\end{eqnarray}
Restricted to ${\cal H}$, our choice of lapse and shift imply
\begin{eqnarray}
   \beta  &=& -\frac{1}{2}\ln{r_{,\lambda}}
   \label{eq:b}
   \\
   U^A &=& -\frac{e^{2\beta}}{r^2} r_{,B} h^{AB}
   \label{eq:U}
   \\
   V &=& -2\, r r_{,u} + \frac{e^{2\beta}}{r} r_{,A} r_{,B} h^{AB}.
   \label{eq:V}
\end{eqnarray}

The structure of the ${\cal J}_u$-hypersurface equations and the evolution
equations~\cite{newt,nullinf} reveals the horizon boundary data necessary for
characteristic evolution. The hypersurface equations are
\begin{eqnarray}
\beta_{,r} &=& \frac{1}{16}rh^{AC}h^{BD}h_{AB,r}h_{CD,r},
\label{eq:beta}
\\
(r^4e^{-2\beta}h_{AB}U^B_{,r})_{,r}  &=&
2r^4 \left(r^{-2}\beta_{,A}\right)_{,r}
-r^2 h^{BC}D_{C}h_{AB,r}
\label{eq:u}
\\
2e^{-2\beta}V_{,r} &=& {\cal R} - 2 D^{A} D_{A} \beta
-2 D^{A}\beta D_{A}\beta + r^{-2} e^{-2\beta} D_{A}(r^4U^A)_{,r} \nonumber \\
&-&\frac{1}{2}r^4e^{-4\beta}h_{AB}U^A_{,r}U^B_{,r} \, ,
\label{eq:v}
\end{eqnarray}
where here $D_A$ is the covariant derivative and ${\cal R}$ the
curvature scalar of the 2-metric $h_{AB}$ of the $r=const$ surfaces
(which differ from the corresponding quantities on the $\lambda=const$
surfaces). These equations form a hierarchy which can be integrated
radially in order to determine $\beta$, $U^A$ and $V$ on a hypersurface
${\cal J}_u$ in terms of integration constants on ${\cal S}_u$, once
the null data $h_{AB}$ has been evolved to ${\cal J}_u$.

The evolution variable $h_{AB}$ can be recast as a single complex function,
since $\det (h_{AB})= \det (q_{AB})=q(x^A)$ is independent of $u$ and $r$. The
code treats such functions on the sphere in terms of stereographic angular
coordinates based upon the auxiliary unit sphere metric $q_{AB}$. Tensor fields
are represented by spin-weighted functions using a computational version of the
$\eth$-formalism~\cite{competh} based upon a complex dyad $q_A$, satisfying
$q_{AB}=q_{(A}\bar q_{B)}$. (Note that this departs from other
conventions~\cite{penrin} in order to avoid unnecessary factors of $\sqrt{2}$
which would be awkward in numerical work.) For example, the vector field $v_A$
is represented by the spin-weight 1 function $v=q^Av_A$. Derivatives of tensor
fields are represented by $\eth$ operators on spin-weighted functions by taking
dyad components of covariant derivatives with respect to the unit sphere
metric.  Our conventions are fixed in the case of a scalar field $\Psi$ by
$\eth \Psi =q^A \nabla_A \Psi$. In these conventions,
\begin{equation}
      (\bar \eth \eth -\eth \bar \eth)\eta=2s\eta
\label{eq:commut}
\end{equation}
for a spin-weight $s$ field $\eta$.

The conformal metric $h_{AB}$ is represented by the dyad components
$J=h_{AB}q^A q^B /2$ and $K=h_{AB}q^A \bar q^B /2$, with the determinant
condition implying $K^2 =1+J\bar J$.   As discussed in~\cite{competh}, the
curvature scalar corresponding to  $h_{AB}$ is
\begin{equation}
  {\cal R} = 2K -\eth \bar \eth K
    + {1\over 2} [\bar\eth^2 J + \eth^2 \bar J]
    +  {1\over 4 K}[\bar \eth \bar J \eth J-\bar \eth J \eth \bar J ] .
\label{eq:rscalar}
\end{equation}

In terms of $J$, the evolution equation takes the form
\begin{equation}
     2 \left(rJ\right)_{,ur}
    - \left(r^{-1}V\left(rJ\right)_{,r}\right)_{,r} = S_J
\label{eq:wev}
\end{equation}
where $S_J$ is explicitly given in term of spin-weighted fields in
Ref.~\cite{high}. $S_J$ contains only first derivatives of $J$ and
predetermined hypersurface quantities so that it does not play a major role in
the integration of Eq.~(\ref{eq:wev}) on a given null hypersurface ${\cal
J}_u$.

Similarly, in integrating Eq.~(\ref{eq:u}) for $U^A$, the code uses the
spin-weight 1 fields $U=q_A U^A$ and $Q=q^A Q_A$, where
\begin{equation}
     Q_A = r^2 e^{-2\,\beta} h_{AB} U^B_{,r},
\end{equation}
so that Eq.~(\ref{eq:u}) reduces to the first order radial equations
\begin{eqnarray}
 \left(r^2 Q  \right)_{,r} &=& q^A[2r^4 \left(r^{-2} \beta_{,A} \right)_{,r}
    -r^2 h^{BC} D_{C} h_{AB,r} ],
   \label{eq:Qa}
    \\
  U_{,r} &=& r^{-2} e^{2\beta} q_A h^{AB}Q_B,
\label{eq:ua}
\end{eqnarray}
with the right-hand sides then rewritten in the $\eth$-formalism in terms of
spin-weighted fields.

The integration constants necessary to evolve Eqs.~(\ref{eq:beta}) -
(\ref{eq:wev}) are the boundary values of $r$, $J$, $\beta$, $Q$, $U$ and $V$
on ${\cal H}$. Part of these boundary conditions are determined by the
characteristic data on ${\cal H}$ and the remainder are determined by gauge
conditions and the solution of the propagation equations. When the horizon
${\cal H}$ has constant $r$ (as in the Schwarzschild case), this is precisely
the data necessary to initiate the radial integrations in the PITT code. In
the generic case, ${\cal H}$ does {\it not} lie on grid points and the
initialization of the $r$-grid boundary also requires $\partial_r J$,
$\partial_r \beta$ and $\partial_r V$ on ${\cal H}$ to provide a Taylor
expansion consistent with a second order accurate code..

\subsection{From horizon variables to Bondi metric functions}
\label{sec:fromto}

We now describe the calculation of the Bondi-Sachs metric variables in a
neighborhood of the horizon as necessary for a characteristic evolution of the
exterior space-time. The essential piece of free horizon data is the conformal
metric $h_{AB}$ or equivalently the metric function $J$. In
Sec.~\ref{sec:confmod}, this data is supplied for the case of a binary
horizon by the conformal model, in which case the value of $r$ on the horizon
is also supplied. In a more general situation, where only $J$ is prescribed,
$r$ can be determined from its value and time derivative $\dot r$ on $S^{-}$
using the ingoing Raychaudhuri equation~(\ref{eq:ruu}), which in spin-weighted
form is
\begin{equation}
      \ddot r =-\frac{r}{4} (\dot J \dot {\bar J}-\dot K^2).
\label{eq:ruus}
\end{equation}
Here, we have used the dyad expansion of the conformal metric $h^{AB}$ which
can be written in terms of $J$ and $K$ as
\begin{equation}
   2\, h^{AB} = - \bar{J} q^A q^B - J \bar{q}^A \bar{q}^B
          + K \left(q^A \bar{q}^B + \bar{q}^A q^B \right).
  \label{eq:hAB}
\end{equation}

Once both the intrinsic geometry $J$ and the radial coordinate $r$ are
known, the metric functions $\beta$, $U$ and $W$ (as well as their radial
derivatives) can be calculated using a hierarchy of horizon propagation
equations [similar to the hierarchy of hypersurface and evolution equations
(\ref{eq:b})-(\ref{eq:wev}) used to propagate fields along a ${\cal J}_u$
null hypersurface].

Next in this hierarchy is the propagation Eq.~(\ref{eq:omegadot}) for the twist
$\omega=q^A \omega_A$ , which has spin-weighted form
\begin{eqnarray}
 (r^2\omega)\dot {}&=& r^2 \eth(\frac{\dot r}{r})
         +\frac{1}{2}[ J\bar \eth(r^2 \dot K)+\bar J \eth(r^2\dot J)
                  -K\bar \eth(r^2\dot J)-K\eth(r^2\dot K)] \nonumber \\
         &+& \frac{r^2}{4} ( \dot {\bar J}\eth J -4 \dot K\eth K
            +2\dot K\bar \eth J +3\dot J \eth \bar J
              -2\dot J \bar \eth K  )
  \label{eq:omegadots}
\end{eqnarray}
and determines $\omega$ given its initial value on the slice $S^{-}$
and the conformal horizon geometry encoded in $J$.

Next, the value of $\beta$ follows from Eq.~(\ref{eq:b}) with $r_{,\lambda}$
determined from Eq.~(\ref{eq:rlam}) in the spin-weighted form
\begin{eqnarray}
   \partial_u\partial_\lambda (r^2) &= &
               \frac{1}{2}\bar \eth [K(\eth \log r +\omega)]
               +\frac{1}{2} \eth [K(\bar \eth \log r +\bar \omega)]
                -\frac{1}{2}\bar \eth [J(\bar \eth \log r +\bar \omega)]
             -\frac{1}{2}\eth [\bar J( \eth \log r + \omega)] \nonumber \\
           &-&\frac{1}{2}{\cal R}-\frac{1}{2}J\bar \omega^2
                 -\frac{1}{2}\bar J \omega^2+K \omega\bar \omega ,
\label{eq:rlams}
\end{eqnarray}
where Eq.~(\ref{eq:rscalar}) supplies the spin-weighted form of the curvature
scalar ${\cal R}$ of the metric $h_{AB}$. This determines $r_{,\lambda}$ on the
entire horizon given its value on the initial slice $S^{-}$.

Next, the Bondi metric functions $U$ and $V$ are evaluated on the horizon 
using Eqs.~(\ref{eq:U}) and (\ref{eq:V}) in the spin-weighted forms
\begin{eqnarray}
   U &=& \frac{e^{2\beta}}{r^2}
       \left(J \bar\eth r - K \eth r\right)
   \label{eq:Uh}
   \\
   V &=& - 2\, r r_{,u} + \frac{e^{2\beta}}{2r}
      \left(- \bar{J} \left(\eth r\right)^2
            - J \left(\bar\eth r\right)^2
            + 2\, K \left(\eth r\right) \left(\bar\eth r\right)\right).
\label{eq:Vh}
\end{eqnarray}

In summary, the Bondi metric functions $r$, $J$, $\beta$, $U$ and $V$ are
determined by the hierarchy of Eqs.~(\ref{eq:ruus}) and (\ref{eq:omegadots}) -
(\ref{eq:Vh}). When the value of $r$ is given, as in the conformal model,
Eq.~(\ref{eq:ruus}) is not needed.

\subsection{Extending the Bondi metric off the horizon}
\label{sec:bondioff}

The evaluation of the Bondi metric functions on radial grid points in the
neighborhood of the horizon requires the $r$-derivatives of $J$, $\beta$, $U$
and $V$. For this purpose we note that, on any given ray, the quantity
$\partial_r F$ is
known once both $\partial_\lambda F$ and $r_{,\lambda}$ are determined, e.g.
\begin{equation}
   \partial_r J = \partial_\lambda J / r_{,\lambda}
                = e^{2\beta} \partial_\lambda J .
\label{eq:Jr}
\end{equation}
We assume below that $r_{,\lambda}$ has already been determined on the horizon
by integrating Eq.~(\ref{eq:rlams}).

We obtain $J_{,r}$ on the horizon (and hence  $K_{,r}$) in terms of
$\partial_\lambda J$, which is determined from its initial value on $S^-$
by integrating  Eq.~(\ref{eq:jlamdot}) in the spin-weighted form
\begin{eqnarray}
    0&=&2r^2\partial_u \partial_\lambda J +2r\dot r \partial_\lambda J
         +2r\dot J \partial_\lambda r
              -Jr^2(\dot {\bar J} \partial_\lambda J
                  +\dot J \partial_\lambda \bar J
                    -2\dot K\partial_\lambda K) \nonumber \\
          &-& (1+K^2)(\eth\omega +\omega^2 -2\omega \eth\log r )
       +\omega (J\bar \eth K-K\bar \eth J)
       + \bar\omega(K\eth J-J \eth K)\nonumber \\
      &-&J \bigg ( J\bar\eth \bar \omega
       -K(\eth\bar\omega +\bar \eth \omega)
      +J\bar\omega^2-2K\omega\bar\omega
          +2K(\omega\bar\eth \log r+\bar\omega\eth \log r)
                -2J\bar\omega\bar\eth \log r \bigg ) .
\label{eq:jlamdots}
\end{eqnarray}

We calculate $\beta_{,r}$ from the Raychaudhuri equation for the outgoing
null geodesics, $R_{\lambda\lambda}=0$, which in Sachs coordinates takes
the form
\begin{equation}
  r_{,\lambda\lambda} = -\frac{r}{8} h^{AC} h^{BD}
     h_{AB,\lambda} h_{CD,\lambda} \, ,
\end{equation}
with spin-weighted version
\begin{equation}
  r_{,\lambda\lambda} = \frac{r}{4} \left(
      \left(K_{,\lambda} \right)^2 - J_{,\lambda}\bar{J}_{,\lambda}
  \right) ,
\end{equation}
which, together with Eq.~(\ref{eq:b}), gives
\begin{equation}
   \beta_{,r}  = \frac{r}{8} (J_{,r} \bar J_{,r} - K_{,r}^2) .
\end{equation}

We obtain $U_{,r}$ (or $Q$) from the twist, which in Bondi coordinates takes
the form
\begin{equation}
    \omega_A = \partial_A \beta +\frac{1}{r}\partial_A r
           +\frac{1}{2}\partial_r ( h_{AB})h^{BC}\partial_C r
            -\frac{r^2}{2}e^{-2\beta}h_{AB}\partial_r U^B,
\label{eq:Ucr}
\end{equation}
with spin-weighted version
\begin{equation}
    \omega = \eth \beta +\frac{1}{r}\eth r
      +\frac{1}{2} (\partial_r J)(K\bar \eth r-\bar J \eth r)
     +\frac{1}{2}(\partial_r K)(K\eth r-J \bar \eth r)
            -\frac{Q}{2}.
\label{eq:twist}
\end{equation}
This determines the boundary data for
$Q$ in terms of quantities that are already known on the horizon.
Given $Q$, the value of $\partial_r U$ follows from Eq.~(\ref{eq:ua}), which
has spin-weighted form
\begin{equation}
   U_{,r} = \frac{e^{2\beta}}{r^2}\left(K Q - J \bar{Q} \right) .
\end{equation}

Finally, we compute $\partial_r V$ by obtaining  $ V_{,\lambda}$ from the
$\lambda$-derivative of Eq.~(\ref{eq:VH}),
\begin{eqnarray}
   \frac{e^{-2\beta}}{r}
   \left[ V_{,\lambda}
          - V \left( 2\,\beta_{,\lambda} + \frac{r_{,\lambda}}{r} \right)
   \right]
 &=&   (r_{,\lambda})^2_{,\lambda} W
     + (r_{,\lambda})^2 W_{,\lambda}
     - 2\, \left(r_{,\lambda} r_{,A} \right)_{,\lambda} W^{A}
     - 2\, r_{,\lambda} r_{,A} W^{A}_{,\lambda} \nonumber \\
 &-&   2\, r_{,\lambda\lambda} r_{,u}
     - 2\, r_{,\lambda} r_{,\lambda u}
     + 2\, \frac{r_{,\lambda A} r_{,B}}{r^2} h^{AB}
     - 2\, \frac{r_{,\lambda} r_{,A} r_{,B}}{r^3} h^{AB} \nonumber \\
 &+& \frac{r_{,A} r_{,B}}{r^2} h^{AB}_{,\lambda} .
\label{eq:vlambda}
\end{eqnarray}
The first three terms in the right-hand side vanish on ${\cal H}$ due to the
gauge conditions. We express the others in spin-weighted form using
Eq.~(\ref{eq:wA}) to obtain
\begin{equation}
  -2\,r_{,A} W^{A}_{,\lambda} = \frac{2}{r^2} \left(
      -\bar{J} \omega \eth r - J \bar{\omega} \bar\eth r
      + K \left( \bar\omega \eth r + \omega \bar\eth r \right)
   \right)
\end{equation}
and Eq.~(\ref{eq:hAB}) to obtain
\begin{eqnarray}
 2\, r_{,\lambda A} r_{,B} h^{AB} &=&
     - \bar{J} (\eth r_{,\lambda}) (\eth r)
     - J (\bar\eth r_{,\lambda}) (\bar\eth r)
     + K \left( (\eth r_{,\lambda}) (\bar \eth r)
     + (\bar\eth r_{,\lambda}) (\eth r) \right)
 \\
   2\, r_{,A} r_{,B} h^{AB}_{,\lambda} &=&
     - \bar{J}_{,\lambda} (\eth r)^2
     - J_{,\lambda} (\bar\eth r)^2
     + 2\, K_{,\lambda} (\eth r) (\bar \eth r) .
\end{eqnarray}
Then Eq.~(\ref{eq:vlambda}) gives
\begin{eqnarray}
  V_{,\lambda} &=& V\left( 2\,\beta_{,\lambda} + \frac{e^{-2\beta}}{r}\right)
               + 4\, r\, r_{,u} \beta_{,\lambda} - 2\, r\, r_{,\lambda u}
 + \frac{2}{r} \left( -\bar{J} \omega \eth r - J \bar{\omega} \bar\eth r
         + K \left( \bar\omega \eth r + \omega \bar\eth r \right) \right)
  \nonumber \\
   &+& \frac{e^{2\beta}}{r} \left(
     - \bar{J} \left(\eth r_{,\lambda}\right) \left(\eth r\right)
     - J \left(\bar\eth r_{,\lambda}\right) \left(\bar\eth r\right)
     + K \left(\bar\eth r_{,\lambda}\right) \left(\eth r\right)
     + K \left(\eth r_{,\lambda}\right) \left(\bar\eth r\right)
   \right)
  \nonumber \\
   &+& \frac{1}{r^2} \left(
       \bar{J} \left(\eth r\right)^2
     + J \left(\bar\eth r\right)^2
     - 2\, K \left(\eth r\right) \left(\bar\eth r\right)
   \right)
  \nonumber \\
   &+& \frac{e^{2\beta}}{2r} \left(
     - \bar{J}_{,\lambda} \left(\eth r\right)^2
     - J_{,\lambda} \left(\bar\eth r\right)^2
     + 2\, K_{,\lambda} \left(\eth r\right) \left(\bar\eth r\right)
   \right)
\end{eqnarray}
in terms of previously determined quantities on the right-hand side.

The values of each metric function $(J,\beta,U,V)$ and its first radial
derivative can then be used to consistently and accurately initialize $r$-grid
points near the horizon. This in turn allows the evolution code to determine
the entire region extending from the horizon to ${\cal I}^+$, as long as the
coordinate system remains well behaved.

\section{Conformal model for the axisymmetric head-on collision}
\label{sec:confmod}

\begin{figure}
\centerline{\epsfxsize=6.0in\leavevmode\epsfbox{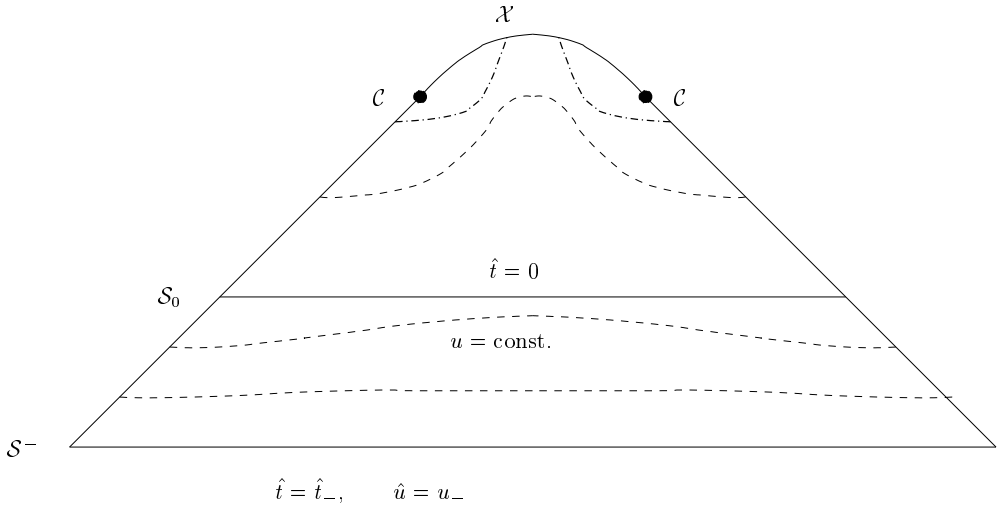}}
\caption{The ingoing null hypersurface normal to an spheroid ${\cal S}_0$ in
Minkowski space pinches off in the future at points ${\cal X}$, where two null
rays cross, bounded by a caustic set ${\cal C}$, where neighboring rays focus;
and it expands in the past to a quasi-spherical slice ${\cal S}^-$.
The foliation by Minkowski time $\hat t$ is is shown by horizontal slices.
The conformal model induces a white hole affine foliation $u$, indicated by
dashed lines, whose upward bulge relative to the Minkowski foliation produces
the fission into disjoint white holes depicted in the final $u$-slice shown. }
\label{fig:spheroid}
\end{figure}

The conformal horizon model~\cite{ndata,asym} supplies the conformal metric
$h_{AB}$ constituting the null data for a binary black hole. The conformal
model is based upon the flat space null hypersurface ${\cal H}$ emanating
normal from a convex, topological sphere ${\cal S}_0$ embedded at a constant
inertial time $\hat t=0$ in Minkowski space. Traced back into the past, ${\cal
H}$ expands to an asymptotically spherical shape. Traced into the future,
${\cal H}$ pinches off where its null rays cross, at points ${\cal X}$, or where
neighboring null rays focus, at caustic points ${\cal C}$. Figure
\ref{fig:spheroid} schematically illustrates the embedding of ${\cal H}$ in
Minkowski space for the case when ${\cal S}_0$ is a prolate spheroid. The
spheoridal case generates the horizon for the axisymmetric head-on collision
of two black holes, or here in the corresponding time reversed scenario, the
horizon for an initially Schwarzschild white hole which undergoes an
axisymmetric fission into two white hole components. The white hole horizon
shares the same submanifold ${\cal H}$ and the same (degenerate) confomal
metric as its Minkowski space counterpart but its surface area and affine
parametrization differ. As a white hole horizon, ${\cal H}$ extends infinitely
far to the past of ${\cal S}_0$ to an asymptotic equilibrium with finite
surface area.

The intrinsic white hole geometry of ${\cal H}$ is obtained by
the following construction based upon the flat space null hypersurface. A $\hat
t$ foliation of ${\cal H}$ is induced by the $\hat t$ foliation of Minkowski
space, with $\hat t$ and the Euclidean coordinates $(x,y,z)$ determining a
Minkowski coordinate system. The prolate spheroid ${\cal S}_0$ is given by
\begin{equation}
           x^2+y^2+\frac{z^2}{1+\epsilon}=a^2 \, ,
\end{equation}
with $\epsilon>0$, or alternatively in angular coordinates
$y^A = (\eta, \phi)$ by
\begin{eqnarray}
   x&=& a \sin\eta \cos\phi \, \\
   y&=& a \sin\eta \sin\phi \, \\
   z&=& \sqrt{1+\epsilon} a \cos\eta \, .
\end{eqnarray}
In these coordinates, the Minkowski metric induces on ${\cal S}_0$ the
intrinsic metric
\begin{equation}
{ \hat g}_{AB} dy^A dy^B =a^2  \left( \left(1 + \epsilon \sin^2 \eta \right)
   d\eta^2 + \sin^2 \eta \, d\phi^2 \right)\, .
\end{equation}

The principal curvature directions on ${\cal S}_0$ are the polar and azimuthal
directions, with corresponding radii of curvature
\begin{equation}
   r_{\eta} = a \frac{(1 + \epsilon \sin^2 \eta )^{3/2}}{ \sqrt{1+\epsilon}}
\end{equation}
and
\begin{equation}
   r_{\phi} = a \frac{(1 + \epsilon \sin^2 \eta )^{1/2}}{\sqrt{1+\epsilon} }
    \, .
\end{equation}
Each generator of ${\cal H}$ encounters two caustics at $\hat t=
r_{\eta}$ and $\hat t= r_{\phi}$ (or one degenerate caustic) if
continued into the future but, along a typical generator, ${\cal H}$ first
pinches off at a cross over with another generator before a caustic is reached.

The time dependent metric of the $\hat t$ foliation is given by
\begin{equation}
     {\hat g}_{AB} dy^A dy^B = a^2 \left(
  \left(1-\frac{\hat t}{r_\eta}\right)^2 (1 + \epsilon \sin^2 \eta) d\eta^2
+ \left(1-\frac{\hat t}{r_\phi}\right)^2               \sin^2 \eta  d\phi^2
                                   \right)  \, .
\label{eq:ghat}
\end{equation}
As $\hat t\rightarrow -\infty$,
$g_{\phi\phi}/g_{\eta\eta}\rightarrow  \sin^2\eta
[1 + \epsilon\sin^2\eta]$ so that the conformal metric in these
coordinates does not approach the standard form of the unit sphere conformal
metric. For this purpose, it is convenient to introduce new angular
coordinates $x^A=(\theta,\phi)$ in which the conformal metric
asymptotes to the unit sphere metric $q_{AB}dx^A dx^B=d\theta^2+\sin^2\theta
d\phi^2$. This requires that
\begin{equation}
   \frac{d\theta}{\sin\theta}= \frac{d\eta}
    {\sin\eta \sqrt {1 +\epsilon \sin^2 \eta }}
\label{eq:angles}
\end{equation}
with the solution
\begin{equation}
   \tan\theta =\sqrt{1+\epsilon}\tan\eta.
\end{equation}
(Here the boost freedom corresponding to the unit sphere conformal group
has been fixed by requiring the transformation to have reflection symmetry
about the equator.)

In these $x^A$ coordinates, $\hat g_{AB}dx^A dx^B= \hat r^2
\hat h_{AB}dx^A dx^B$, where $\det(\hat h_{AB})=\det (q_{AB})=q$,
\begin{equation}
  \hat r^2 =  a^2 \frac{1+\epsilon}{(1 + \epsilon \cos^2 \theta)^2}
  \left(1-\frac{\hat t}{r_\theta}\right)\left(1-\frac{\hat t}{r_\phi}\right)
\label{eq:def_rhat}
\end{equation}
\begin{equation}
  r_{\theta} = r_{\eta}= a \frac{1+\epsilon}{(1 + \epsilon\cos^2 \theta)^{3/2}}
\end{equation}
and
\begin{equation}
     r_{\phi} = \frac{a}{ (1 + \epsilon \cos^2 \theta)^{1/2}} \, .
\label{eq:rphi}
\end{equation}
In the prolate case ($\epsilon > 0$), $ r_{\theta} >  r_{\phi}$.

We apply the conformal horizon model to endow ${\cal H}$ with the intrinsic
metric $g_{AB}=\Omega^2 \hat g_{AB}$ of a white hole horizon. The
conformal factor  $\Omega$ is designed to stop the expansion of the white hole
in the past so that the surface area asymptotically hovers at a fixed radius.
As a result, $h_{AB}(\hat t, x^A)=\hat h_{AB}(\hat t, x^A)$ is the intrinsic
conformal metric of the horizon (as well as of the flat space null
hypersurface). With the dyad choice $q^A=(1,i/\sin\theta)$, Eq.~(\ref{eq:ghat})
implies that the spin-weight-2 field
\begin{equation}
  J={1\over 2}q^A q^B h_{AB}(\hat t,x^C)
      =\frac{1}{2}\bigg ( \frac { \hat t -r_\theta}
                       {\hat t -r_\phi} \bigg )
                    -\frac{1}{2}\bigg ( \frac {\hat t -r_\phi}
                       { \hat t -r_\theta} \bigg )
\label{eq:headonj}
\end{equation}
is the conformal null data for the white hole horizon in the $\hat t$
foliation.

The surface area of the white hole is related to the corresponding surface area
of the flat space null hypersurface by  $r=\Omega \hat r$. A conformal factor
with all required behavior to produce a non-singular white hole is given
by~\cite{ndata,asym}
\begin{equation}
    \Omega=-R_{\infty}\big( \hat u
          +\frac{\sigma^2}{12(p-\hat u )}\big)^{-1},
\label{eq:ansatz}
\end{equation}
where $R_{\infty}$ is the initial equilibrium radius, $p$ is a model parameter
(designated by $\rho$ in Refs.~\cite{ndata,asym}), $\sigma$ is the difference
between the principal curvature radii,
\begin{equation}
\sigma = |r_{\theta} - r_{\phi}| =
       \frac {a |\epsilon| \sin^2 \theta}
         {(1 + \epsilon \cos^2 \theta )^{3/2}} ,
\label{eq:sigma}
\end{equation}
and $\hat u$ is an affine parameter along the generators of ${\cal H}$ with
the same scale as $\hat t$ but with origin $\hat u =0$ chosen to lie midway
between the caustics, i.e. $\hat u =\hat t -r_0$ where
\begin{equation}
  r_0 =\frac {(r_{\theta} + r_{\phi})}{2} =
       a \frac{2 + 2\epsilon - \epsilon\sin^2 \theta}
      {2(1 + \epsilon \cos^2 \theta )^{3/2}}.
\label{eq:r0}
\end{equation}
is the mean curvature of ${\cal S}_0$. For an initially Schwarzschild white
hole, $R_{\infty}=2M$.

As a 3-manifold with boundary, ${\cal H}$ represents both the white hole
horizon and the flat space null hypersurface. Both extend infinitely into the
past and continue into the future to the boundary where ${\cal H}$ pinches
off at caustics and cross over points. Smoothness of the white hole requires
that the parameter $p \ge \sigma_M/\sqrt{13}$, where $\sigma_M$ is the
maximum value of $\sigma$ attained on ${\cal S}_0$. For a prolate spheroid,
the maximum occurs at the equator and $\sigma_M = a |\epsilon|$.

${\cal H}$ must obey the Raychaudhuri equation (\ref{eq:ruu}) both as a flat
space null hypersurface and as a white hole horizon. Both geometries have the
same intrinsic conformal metric so that they must have the same rate of
focusing in terms of their respective affine parameters, in accord with the
Sachs optical equations~\cite{sachseq}.  As a result of the different behavior
of their surface areas due the conformal factor $\Omega$, the affine parameter
$t$ on the white hole horizon is related to its flat space counterpart $\hat t$
according to~\cite{ndata}
\begin{equation}
      \frac {dt}{d \hat t}= \Lambda^{-1}
            =\frac{9}{(12 \hat u (\hat u-p) -
      \sigma^2)^2} \frac{( 5 p + \mu-2 \hat u)^{2 \,
          (2p/\mu +1) \,}}{( 5 p- \mu-2 \hat u)^{2 \,
          (2p/\mu -1) \,} } \, ,
\label{eq:lampr}
\end{equation}
where
\begin{equation}
     \mu = \sqrt{13p^2 -\sigma^2}.
\label{eq:def_mu}
\end{equation}
Here the affine scale of $t$ is fixed by the condition $dt/ d\hat t
\rightarrow 1$ as $\hat t \rightarrow -\infty $. Equation~(\ref{eq:lampr})
determines the deviation of a slicing adapted to an affine parameter
of the white hole horizon from the original slicing given by the Minkowski
embedding. The angular dependence of the
crossover time $t_{\cal X}$ at which the horizon pinches off leads to the change in
topology of the white hole associated with a pair-of-pants shaped horizon.
{}From Eq.~(\ref{eq:rphi}), the pinch-off occurs at $\hat t_{\cal X}
=r_\phi=a/\sqrt{1+\epsilon \cos^2 \theta}$, or
\begin{equation}
    \hat u_{\cal X} = -\frac {a\epsilon \sin^2 \theta}
                      {2(1+\epsilon \cos^2 \theta)^{3/2}}.
\end{equation}
Consequently, $\hat u_{\cal X}$ as well as $\sigma$ vanish at both poles so that
integration of Eq.~(\ref{eq:lampr}) implies that $t_{\cal X} \rightarrow
\infty$ at both poles. This creates the two legs of the pair-of-pants since
$t_{\cal X}$ remains finite at all other angles.

Note that $\Lambda = \Lambda(\hat u/a, p/a, \epsilon)$ within our conformal
model, so that $t = t(\hat t/a, p/a, \epsilon)$. The same scale dependence on
$a$ also holds for the horizon variables  $\hat r$ and $J$, as is evident from
Eqs.~(\ref{eq:def_rhat}) and (\ref{eq:headonj}). In addition
Eq.~(\ref{eq:ansatz}) implies $r=r(R_\infty /a,\hat u/a, p/a, \epsilon)$. Thus
we can scale $a=1$ without any loss of generality of the model.

\section{The close approximation and data on ${\cal S}^{-}$}
\label{sec:headon}

In addition to the conformal metric on ${\cal H}$ (discussed in the last
section), specification of the radius and extrinsic curvature of  the initial
slice ${\cal S}^{-}$ completes the necessary data on ${\cal H}$. Our strategy
is to locate ${\cal S}^{-}$ at an early quasi-stationary era and approximate
these data by their equilibrium values as a Schwarzschild white hole. In the
linearized approximation, the conformal metric of ${\cal H}$ corresponds to a
perturbation of the Schwarzschild background. However, a comparison of how the
fully nonlinear data for the head-on collision deviate from their linearized
counterparts reveals features which can not be described perturbatively. The
chief geometrical issues to be discussed here are the asymptotic properties of
the horizon at $I^-$, the behavior where it pinches off and the analogue of the
bifurcation sphere occurring in a Schwarzschild horizon. As indicated in Fig.
\ref{fig:spheroid}, we choose  ${\cal S}^-$ to correspond to an early Minkowski
time $\hat t =\hat t_-$ so that it is initially quasi-spherical for all
eccentricities $\epsilon$ of the ellipsoid ${\cal S}_0$ located at $\hat t =0$.
The criterion that the initialization be at an early time is $|\hat
t_-|>>r_0$.\ In that approximation $dt/d\hat t\approx 1$ on ${\cal S}^-$ and
Eq.~(\ref{eq:headonj}) gives
\begin{equation}
          J_-\approx \frac{ r_\phi -r_\theta}{\hat t_-  }.
\end{equation}

\subsection{Perturbing a Schwarzschild horizon and the close approximation}
\label{sec:perturbations}

In the Cauchy treatment of the close approximation to the axisymmetric head-on
collision of black holes~\cite{pp}, the background spacetime is the exterior
Kruskal quadrant of the extended Schwarzschild space-time. The trousers shape
of the binary horizon is beyond the scope of a perturbative treatment. Here we
present a fully nonlinear characteristic treatment of the (time reversed)
head-on collision as a 1-parameter sequence of binary collisions with the
Schwarzschild case as a limit. A trousers-shaped horizon exists for each
non-Schwarzschild member of the sequence so that it is possible to investigate
its behavior in the  characteristic version of the close approximation.

The conformal horizon model of a Schwarzschild horizon is the highly degenerate
case in which ${\cal S}_0$ is geometrically a sphere.  Then
$\hat h_{ab}=q_{ab}$, $J=0$ and $r=R_{\infty}=2M$, so that the horizon is
stationary. In the $(u,\lambda,x^A)$ coordinates of the Sachs metric, the
Schwarzschild geometry is described by Eq.~(\ref{eq:amet}) with $W^A =0$, $r=2M
-\lambda u/(4M)$ and
\begin{equation}
         W=\frac{2\lambda^2}{\lambda u -8M^2}.
\end{equation}
These coordinates cover the entire Kruskal manifold with remarkably simple
analytic behavior, as observed by Israel~\cite{israel}. On the white hole
horizon, given by $\lambda =0$, $u$ is an affine parameter with its origin
fixed so that $u=0$ on the $r=2M$ bifurcation sphere where $\partial_\lambda r
=0$.

The spin-weighted versions (\ref{eq:ruus}) - (\ref{eq:jlamdots}) of the
Einstein Eqs.~(\ref{eq:ruu}) - (\ref{eq:jlamdot}) lead to the following
linearized equations governing the perturbation of a Schwarzschild horizon. The
ingoing Raychaudhuri Eq.~(\ref{eq:ruus}) simplifies to
\begin{equation}
        \ddot r = 0,
\end{equation}
so that we can set $r=2M$ on $\cal H$, where $M$ is the background
Schwarzschild mass. Equation~(\ref{eq:omegadots}) then reduces to
\begin{equation}
       (r^2\omega)\dot {}= -\frac{1}{2}r^2\bar \eth \dot J,
\end{equation}
so that
\begin{equation}
 \omega=-\bar \eth J/2,
\label{eq:pertomega}
\end{equation}
 where we  fix the constant of
integration by requiring that the perturbed black hole come to equilibrium
as a Schwarzschild black hole with $J=\omega=0$ as $u\rightarrow-\infty$.
Next Eqs.~(\ref{eq:rlams}) and (\ref{eq:jlamdots}) reduce to
\begin{equation}
       r\partial_u\partial_\lambda r =  -\frac{1}{2}
-\frac{1}{4}(\bar\eth^2 J +\eth^2\bar J)
\label{eq:rlamdp}
\end{equation}
and
\begin{equation}
   0=2r\partial_u \partial_\lambda(rJ)+\eth\bar\eth J + J.
\label{eq:jlamdp}
\end{equation}

It is convenient to set $r=\rho \, r_M$, where
\begin{equation}
                     r_M = 2M -\frac {\lambda u}{4M}
\label{eq:rm}
\end{equation}
is the background Schwarzschild value and where, restricted to ${\cal H}$,
$\rho
=1+O(J^2)$. Then, to first order in the perturbation, Eq.
(\ref{eq:rlamdp}) reduces to
\begin{equation}
     r^2\partial_u \partial_\lambda \rho
           =-\frac{1}{4}(\bar\eth^2 J +\eth^2\bar J)
\label{eq:rhodot}
\end{equation}
and Eq.~(\ref{eq:jlamdp}) reduces to
\begin{equation}
    0=2r^2\partial_u \partial_\lambda J+\eth\bar\eth J
        -u\dot J.
\label{eq:jlamdp2}
\end{equation}
Note that in the perturbative limit the right-hand sides of of the
equations for the extrinsic quantities $\omega$, $r_{,\lambda}$ and
$J_{,\lambda}$ only depend on the intrinsic geometry.

As the parameter $\epsilon$ describing the eccentricity of the spheroid
approaches zero, the binary black hole horizon approaches the Schwarzschild
horizon of a single spherically symmetric black hole. Thus the close
approximation for a binary black hole can be described by a perturbation
expansion in $\epsilon$. In order to integrate the perturbation
Eqs.~(\ref{eq:rhodot}) and (\ref{eq:jlamdp2}), with $J$ supplied by the
conformal model, we must relate the affine parameter $t$ of the conformal
model to the restriction to ${\cal H}$ of the coordinate $u$ of the Sachs
metric. We set $t=Au$, where
\begin{equation}
   A = \frac {1} {(1 + \epsilon\cos^2\theta)^{3/2}} =1+O(\epsilon)
\label{eq:capa}
\end{equation}
fixes the relative scale freedom in the affine parameters so that the
perturbation has the early time behavior
\begin{equation}
            J\sim -\frac{a \epsilon\sin^2\theta}{u}
\end{equation}
in accord with the quadrupole nature of the close approximation.

\subsection {Data on ${\cal S}^{-}$}
\label{sec:sminus}

The conformal model described in Sec.~\ref{sec:confmod} supplies the values of
$J$ and $r$ on the entire horizon. Other quantities have to be initiated at an
early cross section  ${\cal S}^-$, near which the horizon behaves as a
perturbed Schwarzschild horizon. As indicated in Fig.~\ref{fig:spheroid}, we
locate ${\cal S}^-$ at a constant Minkowski time $\hat t=\hat t_-$, as well as
at a constant white hole affine time $u=u_-$.  The relationship $\hat t(u)$ is
then determined by integrating $du/d\hat t = 1/(A \Lambda)$ with initial
conditions determined on ${\cal S}^-$ and $\Lambda$ and $A$ given in
Eqs.~(\ref{eq:lampr}) and (\ref{eq:capa}). This determines the horizon data
$J(u,x^A)$ from $J(\hat t,x^A)$ given in Eq.~(\ref{eq:headonj}). Similarly,
$r(u,x^A)$ is determined from $r(\hat t,x^A)=\Omega (\hat t,x^A) \hat r (\hat
t,x^A)$, with $\hat r$ and $\Omega$ given by Eqs.~(\ref{eq:def_rhat}) and
(\ref{eq:ansatz}), respectively. The requirement that ${\cal S}^-$ be located
in the quasi-equilibrium era implies that $r_- \approx 2M$ to an excellent
approximation.

The way in which the values of $u_-$ and $\hat t_-$ determine the location of
${\cal S}_0$ relative to ${\cal S}^-$ depends upon the parameters entering the
conformal model. Even in the limit $\epsilon \rightarrow0$, where the conformal
model yields a Schwarzschild horizon, this relation depends upon the model
parameter $p$. The simplest limiting case is when $p$ and $\epsilon$ both
vanish. Then $A=\Lambda =1$  so that  $\hat t- \hat t_-=u-u_-$ and the
Minkowski spheroid ${\cal S}_0$ (where $\hat t=0$) is located at $u=u_- -\hat
t_-$.    Thus a negative value of $\hat t_-$ locates ${\cal S}_0$ to the future
of ${\cal S^-}$.

In the Schwarzschild limit, $\partial_\lambda r \rightarrow \partial_\lambda
r_M = -u/4M$, where $u=0$ on the $r=2M$ bifurcation sphere ${\cal B}$. This
allows us to use the initial value of $\partial_\lambda r$ to  determine the
location of ${\cal S}^-$ on the horizon by setting
\begin{equation}
     \partial_\lambda r_-  = - \frac{u_-}{4M}
\label{eq:uinit}
\end{equation}
on ${\cal S}^-$. Thus specification of the initial outward expansion of ${\cal
S}^-$ fixes the translation freedom in the affine parameter $u$. The
requirement $|u_-| >>M$ ensures that the initialization be at an early time.

A 1-parameter family of horizon data for a head-on collision results from
choosing ${\cal S}_0$ to be an $\epsilon$-family of spheroids (with
$\epsilon\ge 0$). The close approximation, corresponding to the behavior linear
in $\epsilon$, provides insight into the asymptotic structure of the horizon
at early times. In this linear approximation, Eq.~(\ref{eq:headonj}) implies
\begin{equation}
     J\approx -\frac {\epsilon a \sin^2\theta}{\hat t -a}
\end{equation}
on the horizon; and Eqs.~(\ref{eq:sigma}) and (\ref{eq:r0}) imply
\begin{equation}
      \sigma \approx \epsilon a\sin^2\theta
\end{equation}
and
\begin{equation}
       r_0 \approx \frac{r_\theta +r_\phi}{2}= a\left(1+\frac{\epsilon}{2}
            -\epsilon \cos^2\theta \right).
\end{equation}
In addition, $r=2M+O(\epsilon^2)$. (We use $\approx$ to denote
approximations valid for small $\epsilon$ and $\sim$ to denote asymptotic
approximations at early times.)

The early time asymptotic behavior of all the horizon variables can be
explicitly evaluated in this approximation. Equation~(\ref{eq:pertomega})
determines the asymptotic dependence
\begin{equation}
     \omega\sim -\bar \eth J/2,
\label{eq:asymomega}
\end{equation}
on the assumption of an initially non-spinning Schwarzschild white hole.
Using the identities $\eth^2\cos^2 \theta=2\sin^2\theta$,
$(\bar\eth^2\eth^2+\eth^2\bar\eth^2)\cos^2\theta =
48(\cos^2\theta-\frac{1}{3})$, the commutation relation~(\ref{eq:commut})
and the property $\bar\eth\eth Y_{\ell m}=-\ell(\ell+1)Y_{\ell m}$, Eq.
(\ref{eq:rhodot}) reduces to
\begin{equation}
     r^2\partial_{\hat t} \partial_\lambda \rho \approx
 \frac{6\epsilon a}{\Lambda (\hat t-a)} \left(\cos^2\theta-\frac{1}{3} \right).
\label{eq:rhoearly}
\end{equation}
At early times, where $\Lambda\sim 1$, this integrates to give
\begin{equation}
      r^2 \partial_\lambda \rho  \sim
  6\epsilon a \left(\cos^2\theta-\frac{1}{3} \right) 
   \log\bigg(\frac{\hat t - a}{\hat t_ - a} \bigg),
\end{equation}
where we set the integration constant so that $\partial_\lambda\rho=0$
on ${\cal S}^-$ in accord with Eq.~(\ref{eq:uinit}).

Similarly, at early times, the integral of Eq.~(\ref{eq:jlamdp2}) gives
\begin{equation}
     \partial_\lambda J \sim -\frac{3\epsilon
       a\sin^2\theta}{2r^2} \log\bigg (\frac{\hat t -a}{\hat t_- -a} \bigg ).
\label{eq:jlamint}
\end{equation}

Note that $\partial_\lambda \rho$ and $\partial_\lambda J$ have  logarithmic
asymptotic behavior at $I^-$. In the exterior evolution code  this singular
behavior is renormalized by dealing instead with the quantities $\partial_r
\rho =\partial_\lambda \rho /\partial_\lambda r$  and $\partial_r
J=\partial_\lambda J/\partial_\lambda r$. Since $\partial_\lambda r \sim
-u/4M$ at $I^-$, these quantities both go to zero as $\log u /u$ as
$u\rightarrow -\infty$.  This justifies initializing these quantities to zero
on ${\cal S^-}$. The initialization error is $O(\log u_-/u_-)$ and converges
to $0$ as $u_-\rightarrow-\infty$.
In summary, given $J$ and $r$ on the horizon via the conformal model, the
remaining data necessary on ${\cal S^-}$ are initialized according to
\begin{eqnarray}
         \omega_- & = & -\frac{1}{2}\bar \eth J_-, \\
\label{eq:omegaminus}
            \partial_\lambda \rho_- & = & 0,  \\
\label{eq:rhominus}
           \partial_\lambda J_- & = & 0.
\label{eq:jminus}
\end{eqnarray}
With this initialization, all Bondi-Sachs start-up variables for the exterior
evolution are asymptotically well defined at $I^-$ except for
$e^{-2\beta}=\partial_\lambda r\sim -u/(4M)$. This is handled by the
renormalization $\beta=\beta_M+\beta_R$ where $e^{-2\beta_M}= -u/(4M)$ and
$\beta_R \rightarrow 0$ at $I^-$. In the evolution code, the singular part
$\beta_M$ is analytically factored out of the Bondi equations. The regular part
is given by
\begin{equation}
       e^{-2\beta_R} \sim 1 - \frac{8M^2 \partial_\lambda \rho}{u}.
\end{equation}
Then, referring to Eq.~(\ref{eq:rhoearly}), $\beta_R$ can be initialized to $0$
on ${\cal S}^-$ with $O(\log u_-/u_-)$ accuracy.

The early time approximation breaks down before reaching the crossover points
${\cal X}$ where $\hat t \approx a$, as evident from Eqs.~(\ref{eq:rhoearly})
and (\ref{eq:jlamint}). The way that the horizon pinches off at ${\cal X}$ in
a sequence of models as $\epsilon\rightarrow 0$ is sensitive to the behavior
of the model parameter $p$ along the sequence. We consider here the case
$p=const$ (independent of $\epsilon$ as well as angle).

The crossover points occur at Minkowski time $\hat t_{\cal X} =  a/\sqrt{1
+\epsilon\cos^2\theta} \approx a [1-(\epsilon/2) \cos^2 \theta ]$. The
corresponding values $u_{\cal X}=t_{\cal X}/A$ are found from integrating Eq.
(\ref{eq:lampr}). The rays on the poles of the prolate spheroid, where $\sigma
=0$, do not focus. Since the horizon surface area is asymptotically constant,
this lack of focusing implies that the horizon persists forever along the
poles, i.e. for $-\infty<u<+\infty$ (independent of the value of $\epsilon$).

In order to investigate the location of ${\cal X}$ along the non-polar rays,
consider the small $\epsilon$ behavior
\begin{equation}
   \frac {1}{\Lambda} \approx \frac {(3p^2 -5p\hat u +\hat u^2)^2}
      {\hat u^2 (\hat u-p)^2}
             \bigg (\frac {(5+\sqrt{13})p -2\hat u}
                 {(5-\sqrt{13})p  -2\hat u}\bigg )^{4/\sqrt{13}},
\end{equation}
where
\begin{equation}
  \hat u\approx \hat t 
    -a \left(1+\frac{\epsilon}{2} -\epsilon \cos^2\theta \right)
       \approx \hat t -\hat t_{\cal X} -\frac{a\epsilon}{2}\sin^2\theta  .
\end{equation}
The dominant behavior near ${\cal X}$ is revealed by the further
approximation
\begin{equation}
        \frac {1}{\Lambda} \approx \frac {9p^2}{\hat u^2}
             \bigg (\frac {5+\sqrt{13}}{5-\sqrt{13}}\bigg )^{4/\sqrt{13}},
\end{equation}
which is valid for $\hat u<<p$, e.g. near ${\cal X}$ where $\hat u\approx
0$. In this approximation,
\begin{equation}
      u_{\cal X}-u_- =\int_{\hat t_-}^{\hat t_{\cal X}}
           \frac {d\hat t}{A\Lambda}  \approx 9p^2
           \bigg (\frac {5+\sqrt{13}}{5-\sqrt{13}}\bigg )^{4/\sqrt{13}}
           \bigg (\frac{2}{a\epsilon\sin^2\theta}
           +\frac {1}{\hat t_- -a}\bigg ),
\label{eq:pinch}
\end{equation}
so that $u_{\cal X}\rightarrow\infty$ as $\epsilon\rightarrow 0$ along all
rays. In this limit, the entire cross over seam on the pair-of-pants is
mapped to $t=\infty$. As a result, the corresponding first order
perturbation theory for this version of the close approximation is
well behaved on the entire white hole horizon, not just the segment bordering
the exterior space-time.

Of special physical importance is the location of the crossover surface ${\cal
X}$ where the horizon pinches off relative to the surface ${\cal B}$ defined by
$\partial_\lambda r=0$. ${\cal B}$ represents a boundary for the Bondi
evolution resulting from the breakdown of the surface area coordinate $r$. In
the $\epsilon =0$ Schwarzschild limit, ${\cal B}$ is the $r=2M$ bifurcation
sphere located at $u=0$ and ${\cal X}$ lies at $u=\infty$ (at $I^+$). In this
limit, the white hole fission takes place in the infinite future. For small but
nonzero $\epsilon$, the dominant $O(1/\epsilon\sin^2\theta)$ dependence in Eq.
(\ref{eq:pinch}), implies that ${\cal X}$ lies at a finite but large value,
with the point at the equator where the white hole fissions (the crotch in the
pair-of-pants picture) located at the earliest point on ${\cal X}$. However,
${\cal B}$ remains within $O(\epsilon)$ of its Schwarzschild location at
$u=0$.  Thus, for small $\epsilon$ the fission is ``hidden'' beyond ${\cal B}$
in the sense that it is not visible to observers at ${\cal I}^+$. {}From the
time reversed view of a black hole merger, the individual black holes would
merge inside a white hole horizon.

\section{Nonlinear horizon data}
\label{sec:mts}

The close approximation results just described for a white hole fission, when
reinterpreted in the time reversed sense of a black hole merger, imply that the
individual black holes merge inside a white hole horizon corresponding to the
marginally anti-trapped branch of the $r=2M$ Schwarzschild surface. Of prime
importance in the non-perturbative regime is whether an entirely different
scenario is possible in which the individual black holes form and merge without
the existence of a marginally anti-trapped surface (MATS) on the event horizon.
The ingoing null hypersurface which intersects the horizon in such a MATS has
an extremum in its surface area. As a result, the Bondi surface area
coordinate based upon an ingoing null foliation is singular at the MATS. A
Bondi evolution carried out backward in time on these ingoing null
hypersurfaces would terminate at the MATS. In particular, the absence of a MATS
to the future of the merger is a necessary condition for a Bondi evolution
backward in time throughout the entire post-merger period. (See the discussion
below for more technical details.) 

It is possible to avoid this problem by means of a null evolution
using an affine parameter rather than a surface area parameter as radial
coordinate. Null evolution codes in different gauges than the Bondi
gauge have been developed~\cite{bartnik1,bartnik2}. However, at present such
codes have not been successful in the stable evolution of dynamic horizons.

Restated from the alternative viewpoint of a white hole fission, as being
pursued here, the absence of a marginally trapped surface (MTS) prior to the
fission is a necessary condition for a Bondi evolution forward in time
throughout the pre-fission period. Thus, the absence of a pre-fission MTS is
necessary in order to carry out our strategy for obtaining the complete
post-merger wave form of a binary black hole by means of a Bondi evolution.

The bifurcation sphere ${\cal B}$ in the Schwarzschild space-time is a MTS on
the white hole horizon (and in this degenerate case also a MATS). As already
discussed in Sec.~\ref{sec:sminus} and  explicitly demonstrated in Sec.
\ref{sec:results}, in the small $\epsilon$ regime of our model the
corresponding white hole fissions at a very late time well beyond the MTS. This
is the expected effect of a non-singular perturbation: in the $\epsilon
\rightarrow 0$ limit there is only one white hole so that the fission is hidden
at $I^+$. The behavior in the non-linear regime is not so easy to predict. The
following discussion of the properties of an MTS and its relation to a Bondi
boundary ${\cal B}$ provides a basis for understanding the computational
results of Sec.~\ref{sec:results}.

It should first be noted that, unlike the definition of a MTS, the definition
of ${\cal B}$ is foliation dependent. A MTS is a topological sphere with one
non-diverging normal null direction and the other divergence-free. The
Bondi boundary ${\cal B}$ on a white hole horizon ${\cal H}$ parameterized by
$u$ is the earliest slice $u=u_{{\cal B}}$ whose outward null normal is not
strictly diverging at all points. In the relation between $(u,\lambda, x^A)$
Sachs coordinates and $(u,r, x^A)$ Bondi coordinates, this implies that
$\partial_\lambda r(u_{{\cal B}},0,x^A)=0$ at some point of ${\cal B}$. In
other foliations of ${\cal H}$, ${\cal B}$ (if it exists) could occur later or
earlier than in the affine foliation considered here.

In the affine foliation, a MTS on of ${\cal H}$ can be described in
Sachs coordinates in the form $u+F(x^A)=0$, $\lambda=0$. The outgoing null
normal to the MTS is
\begin{equation}
       L_a = -\alpha \partial_a \lambda - \partial_a (u+F)
\end{equation}
where
\begin{equation}
       \alpha =\frac{1}{2 r^2} h^{AB} (\partial_A F )\partial_B F
\end{equation}
which, with $n^a \partial_a =\partial_u$, defines the projection tensor
\begin{equation}
       \gamma^{ab} = g^{ab} + 2 L^{(a} n^{b)}.
\end{equation}
The MTS satisfies
\begin{equation}
       \gamma^{ab} \nabla_a L_b =0
\end{equation}
which has the coordinate form
\begin{eqnarray}
       \gamma^{ab} \nabla_a L_b   &=& -\frac{1}{r^2}[D^A D_A F\nonumber \\
     &+&(D^A F)\partial_\lambda g_{uA} +\alpha\partial _u (r^2)
            -\partial_\lambda (r^2)
           -(\partial_u h^{AB}) (D_A F)D_B F]|_{u=-F} \nonumber \\
       &=& 0
\label{eq:mts}
\end{eqnarray}
(where $D_A h_{BC} =0$). The MTS, if it exists, can be located by solving
Eq.~(\ref{eq:mts}).

The following two propositions relate the Bondi coordinates to the
existence of a MTS on a white hole horizon ${\cal H}$:

{\it Proposition I}. A Bondi cross section $u=const$ satisfying
$\partial_\lambda
r =0$ is a MTS.

{\it Proposition II}. A MTS cannot exist in a region $u<u_{{\cal B}}$
in which $\partial_\lambda r > 0$.

The first proposition follows immediately from setting $F=0$ in Eq.
(\ref{eq:mts}). The second proposition follows from noting that at some point
on
the MTS the function $F$ would have a maximum where $D_A F =0$ and where Eq.
(\ref{eq:mts}) would reduce to
 \begin{equation}
       D^A D_A F =  \partial_\lambda (r^2) >0.
           \label{eq:nmts}
\end{equation}
But the inequality in Eq.~(\ref{eq:nmts}) precludes the existence of a maximum.

The second proposition establishes that a MTS cannot form before the Bondi
boundary. Thus a Bondi evolution might terminate prematurely due to an
injudicious choice of foliation of ${\cal H}$.  A computational module for
locating a MTS on a null hypersurface has been developed and successfully used
for long term tracking of  a moving black hole~\cite{wobb}. In future work,
this module will be applied to binary horizons. Here we consider only the less
geometrical Bondi boundary ${\cal B}$.

As $\epsilon$ increases into the nonlinear region, the effect on ${\cal B}$ can
be seen from integrating Eq.~(\ref{eq:rlam}) over the sphere. In doing so, we
note that the terms which are divergences integrate to zero and we can apply
the Gauss-Bonnet theorem to the curvature scalar term to obtain
\begin{equation}
   \partial_u \oint \partial_\lambda (r^2) dS=
           -4\pi +\oint h^{AB}\omega_A\omega_B  dS.
\label{eq:intrlams}
\end{equation}
The term $-4\pi$ is responsible for the formation of ${\cal B}$ in the
background Schwarzschild case. The nonlinear correction due to the  twist is of
an opposite sign and delays the formation. Thus, as $\epsilon$ increases from
$0$, the formation of ${\cal B}$ is delayed while  the location of ${\cal X}$
moves to earlier times. Although Eq.~(\ref{eq:intrlams}) only describes
averaged angular behavior, it suggests that sufficient nonlinearity might cause
the white hole fission, located on the equator of ${\cal X}$, to occur prior to
${\cal B}$ . (In the time reversed case, this would alow the merger of
individual black holes without the necessity of a MATS and the consequent
singularity in its past implied by Penrose's theorem~\cite{pensing}.)

\section{Numerical Results}
\label{sec:results}

The scenario hypothesized at the end of Sec.~\ref{sec:mts} can indeed be
demonstrated by integrating the equations underlying the conformal horizon
model. At an
early time, the equilibrium conditions on the white hole horizon imply that
$r=2M$ and $\partial_\lambda r= -u/4M >0$. As the horizon evolves, the surface
area $r$ decreases along all rays but, for the axisymmetric and  reflection
symmetric fission considered here, it decreases  fastest along the equatorial
rays where the pinch-off first occurs. The outward expansion measured by
$\Theta_{OUT}=2\partial_\lambda r /r$ also initially decreases along all rays,
although this process can be reversed by the growth of nonlinear terms, as
indicated by the ray-averaged behavior governed by Eq.~(\ref{eq:intrlams}). In
the close approximation, the expansion goes to zero along all rays before the
horizon pinches off, i.e. the crotch at the center of the pair of pants is
hidden behind a MTS. The crucial question in the nonlinear regime is whether
the horizon can pinch off before the formation of a MTS, i.e. whether the
crotch is bare.  For the related question in terms of a Bondi boundary rather
than a MTS, the issue is who wins the race toward 0, the radius $r$ or the
expansion $\Theta_{OUT}$ along some ray.

We conduct this race for each of a sequence of models in the range $0 \le
\epsilon \le 10^{-2}$, with the remaining parameters fixed at $M=100$,
$u_-=-100$, $\hat t_- =-10$, $a=1$ and $p=(10^{-2}+10^{-5})/\sqrt{13}$ (just
above the minimum value of $p$ allowed by regularity of the conformal model for
this range of eccentricities).

\begin{figure}
\centerline{ \vbox{\hbox{
\psfig{file=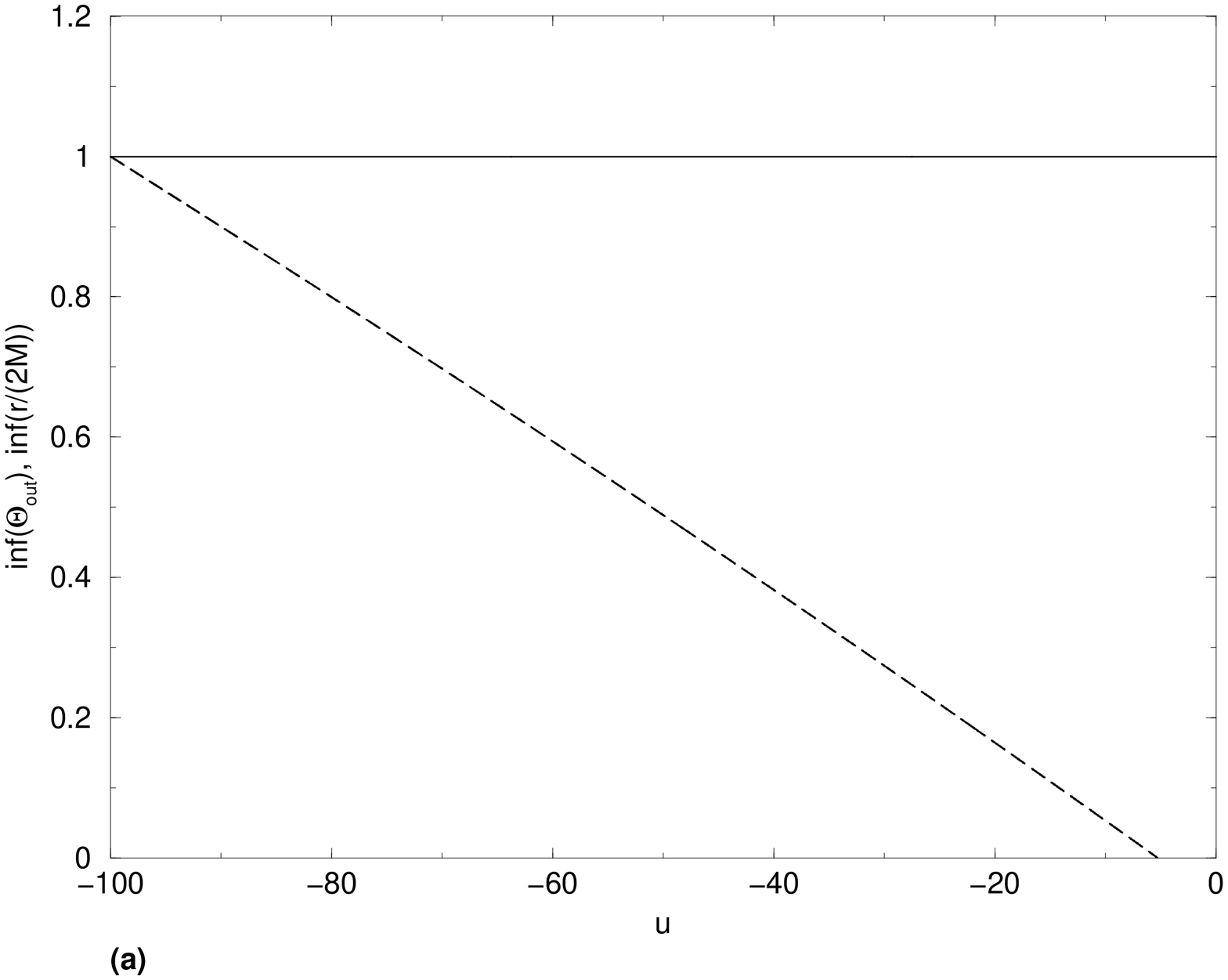,height=2.4in,width=3.5in,angle=-90}}
\hbox{
\psfig{file=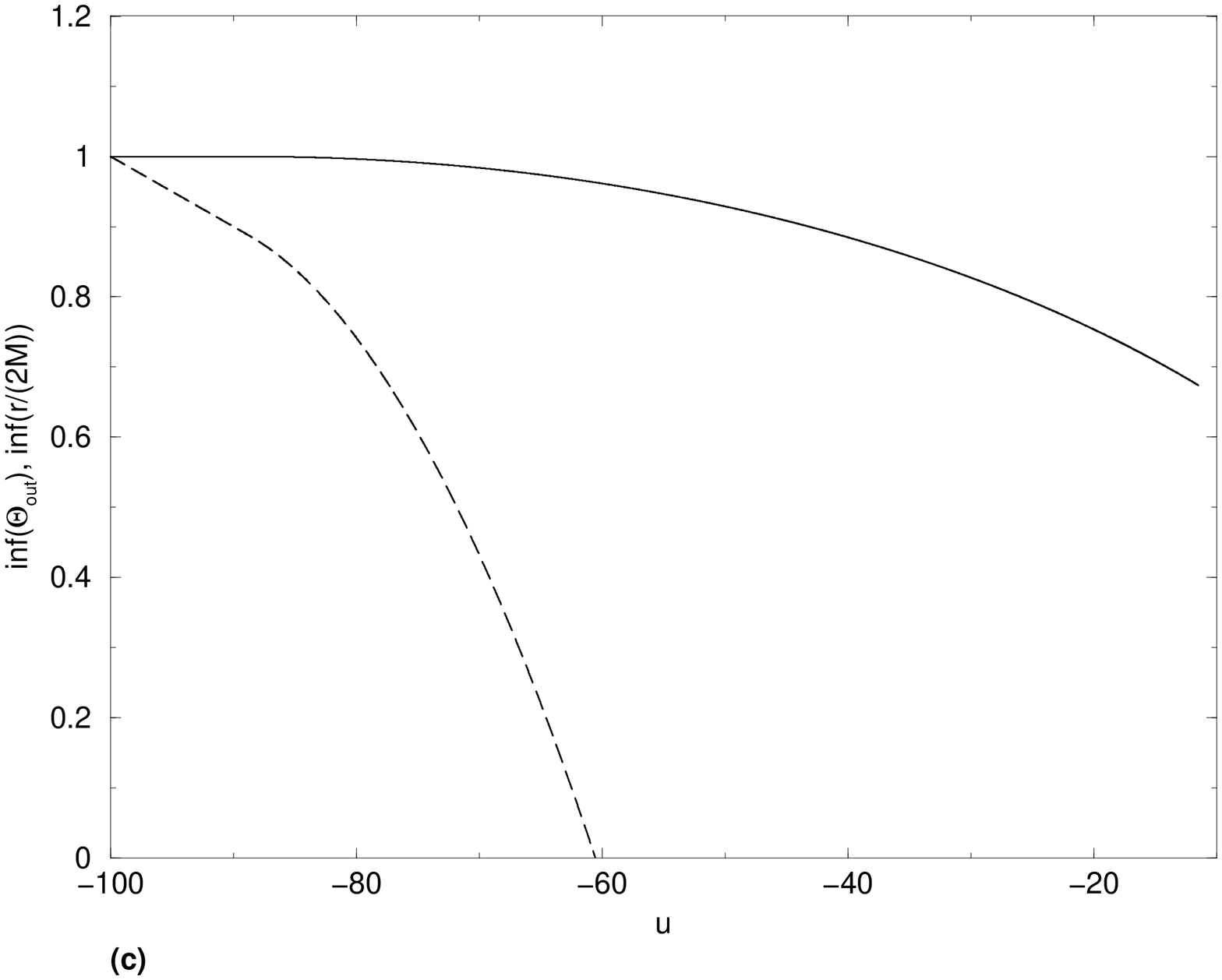,height=2.4in,width=3.5in,angle=-90}}
\hbox{
\psfig{file=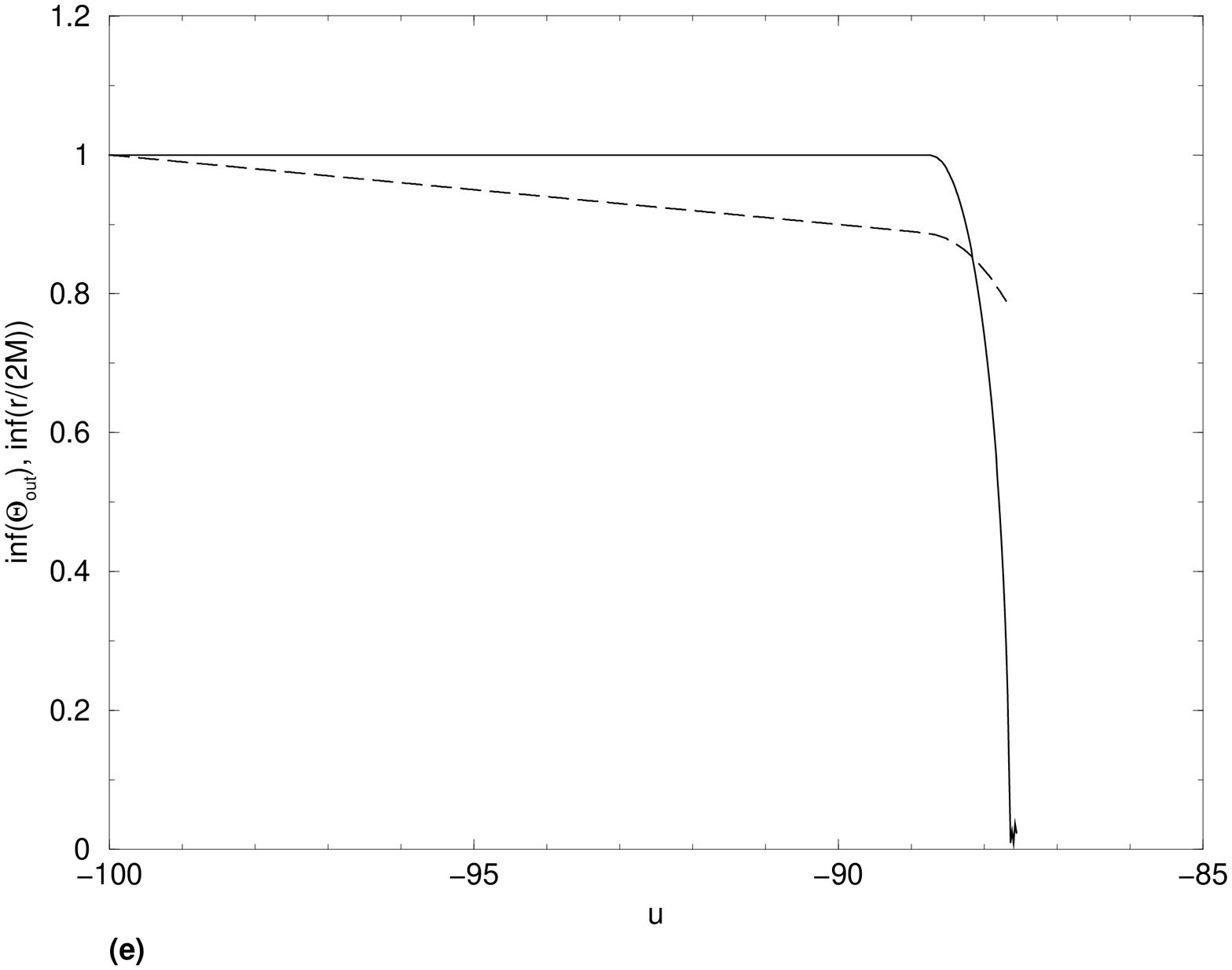,height=2.4in,width=3.5in,angle=-90}}}
\vbox{\hbox{
\psfig{file=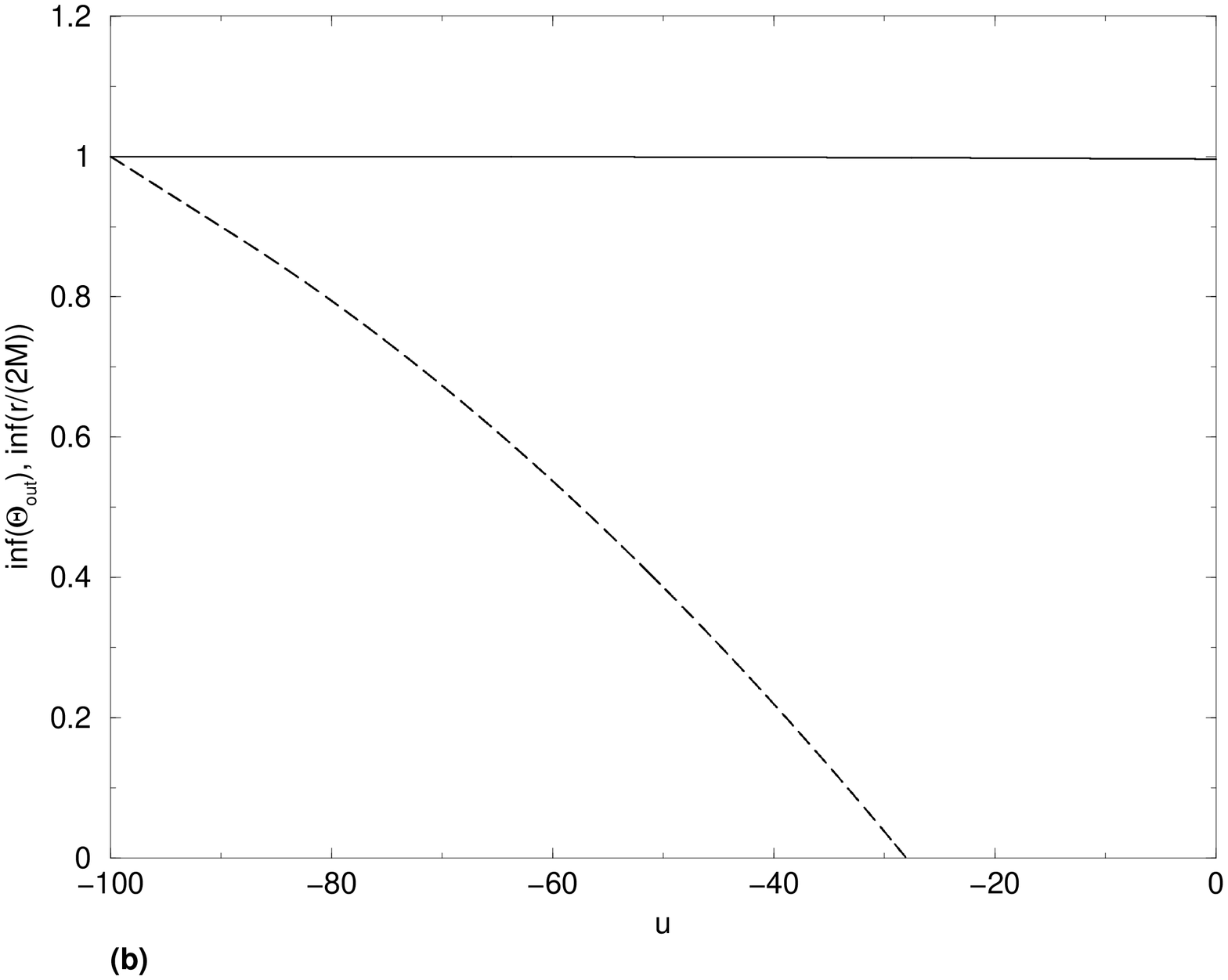,height=2.4in,width=3.5in,angle=-90}}
\hbox{
\psfig{file=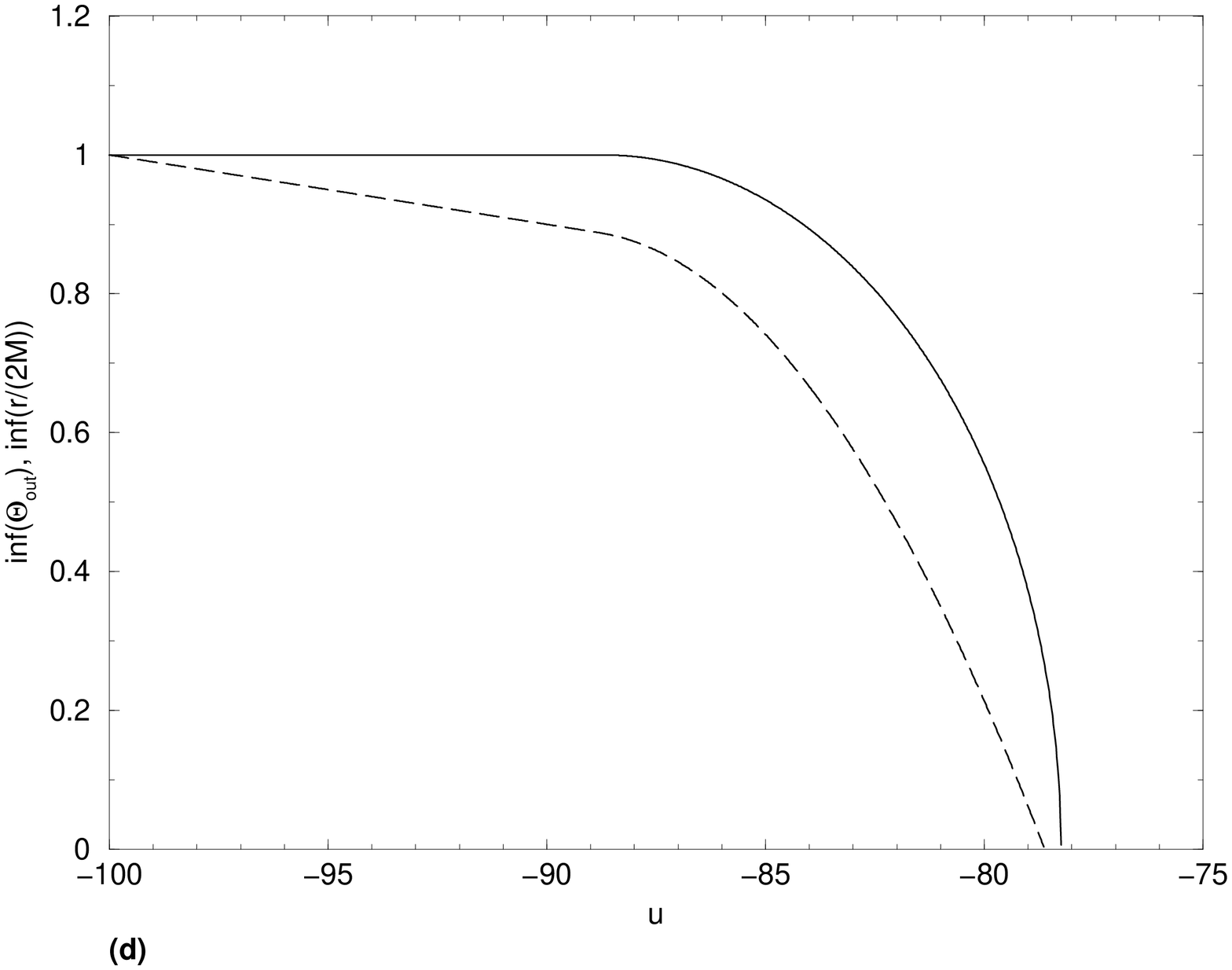,height=2.4in,width=3.5in,angle=-90}}
\hbox{
\psfig{file=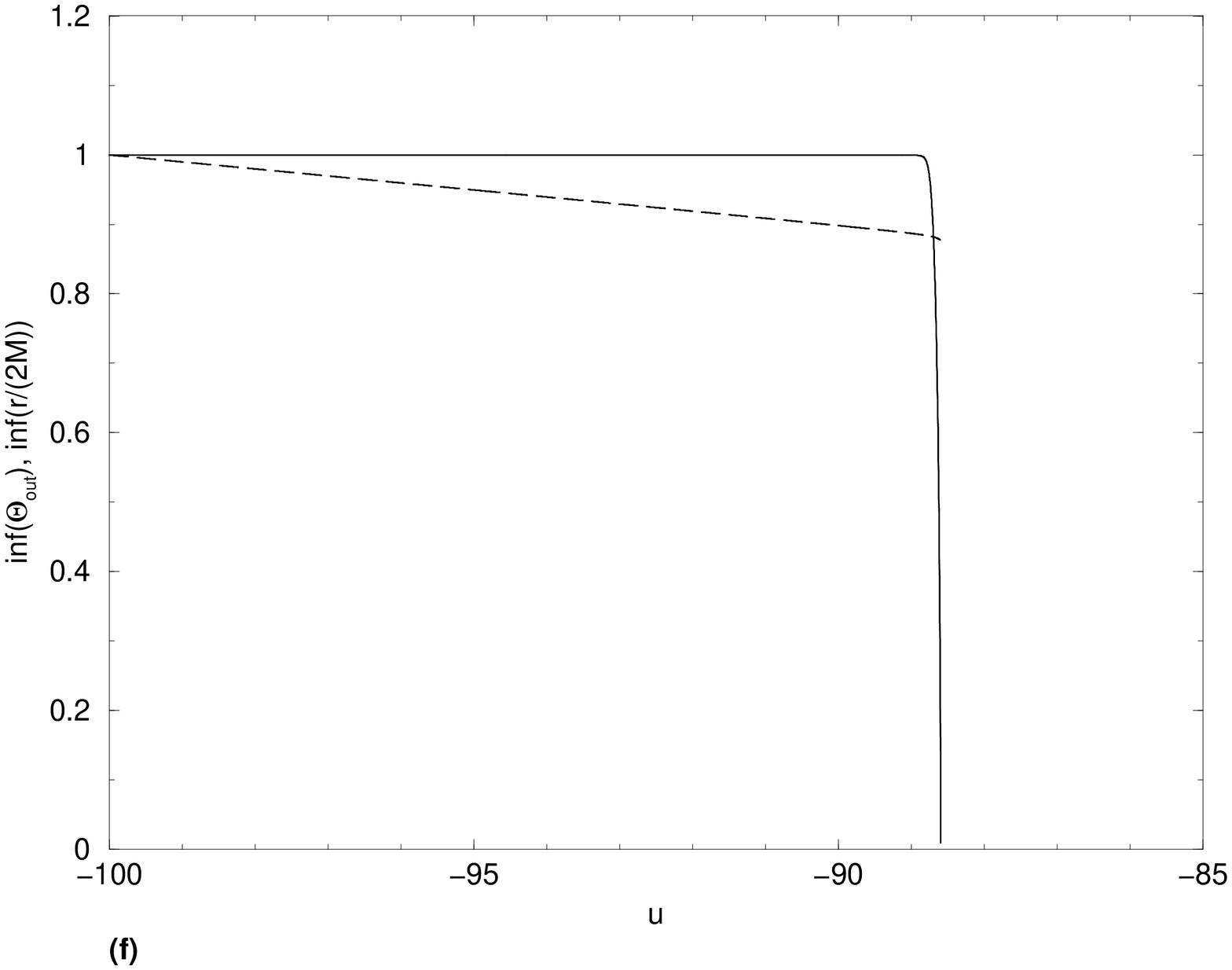,height=2.4in,width=3.5in,angle=-90}}}
} 

\caption{The ``race'' between the minimum values (over the sphere of
rays) of the outward expansion and the radius of the horizon, for
increasing eccentricities $\epsilon=10^{-7}$ (a), $\epsilon=10^{-6}$
(b), $\epsilon=10^{-5}$ (c), $\epsilon=10^{-4}$ (d), $\epsilon=10^{-3}$
(e), $\epsilon=10^{-2}$ (f), all for mass $M=100$ and $t_0=-10$. The
expansion is plotted in terms of $\inf(\Theta_{OUT})$ with dashed lines
and the radius in terms of $\inf\left(r/(2 M)\right)$ with solid lines,
both rescaled to unity at the starting time. For small $\epsilon$, the
radius hovers near $2M$ longer than it takes the Bondi boundary to form
and the fission remains hidden.  For larger values of $\epsilon$, the
fission occurs before the Bondi boundary, as the sequence shows.}
\label{fig:exp-r-race}
\end{figure}

We monitor the minimum value over the sphere of the expansion of the outgoing
null rays on the horizon, and of the Bondi radius of the horizon.  The results
are displayed in Fig.~\ref{fig:exp-r-race} in terms of values of $r$ and
$\Theta_{OUT}$ normalized to 1 at the initial time, so that the race
starts out even.

Panel $(a)$ of the figure, for the small value $\epsilon=10^{-7}$, shows little
deviation from a Schwarzschild horizon. A Bondi boundary ${\cal B}$ forms at
$u\approx 0$ as a consequence of the  zeroth order in $\epsilon$ term ${\cal
R}\approx 2$ in Eq.~(\ref{eq:rlam}) (${\cal R}= 2$ for a unit sphere), which
causes $\Theta_{OUT}$ to decrease linearly with $u$. We have verified that the
initial slope of $\Theta_{OUT}$ as seen in the graph corresponds to the
expected Schwarzschild value.

For $\epsilon=10^{-6}$, still near the close limit, the radius of the horizon
hardly changes before the Bondi horizon forms, as illustrated in panel $(b)$.
However, the deviation of the expansion from a pure linear-in-time behavior is
noticeable and its deviation from spherical symmetry as a function of ray is
also noticeable in the full numerical data. (The angular behavior of the
relevant geometrical quantities is discussed more fully below.)
As $\epsilon$ increases to $10^{-5}$ in panel $(c)$, both the radius and
the expansion show markedly nonlinear behavior but the expansion still readily
wins the race toward zero. However, its margin of
victory gets smaller with increasing $\epsilon$ as manifest in panel $(d)$
for $\epsilon=10^{-4}$ in which the race is nearly a tie.

For $\epsilon=10^{-3}$, as shown in panel $(e)$, the radius now wins the race.
The fission takes place while the expansion is still significantly large, at
about $90\%$ of its initial value. For a larger value of $\epsilon$, as shown
in panel $(f )$, the effect is even more dramatic. The radius makes a
sudden plunge to zero to win the race before the expansion has undergone any
appreciable change. This is a dramatic nonlinear effect. Although the radius
and expansion begin the race at the same starting point (in rescaled units),
the expansion begins with a flying start (its initial slope in the figures) and
gets accelerated by linear effects whereas the radius starts from rest and
only gets accelerated by quadratic or higher nonlinearities.

\begin{figure}
\centerline{\epsfxsize=6in\epsfbox{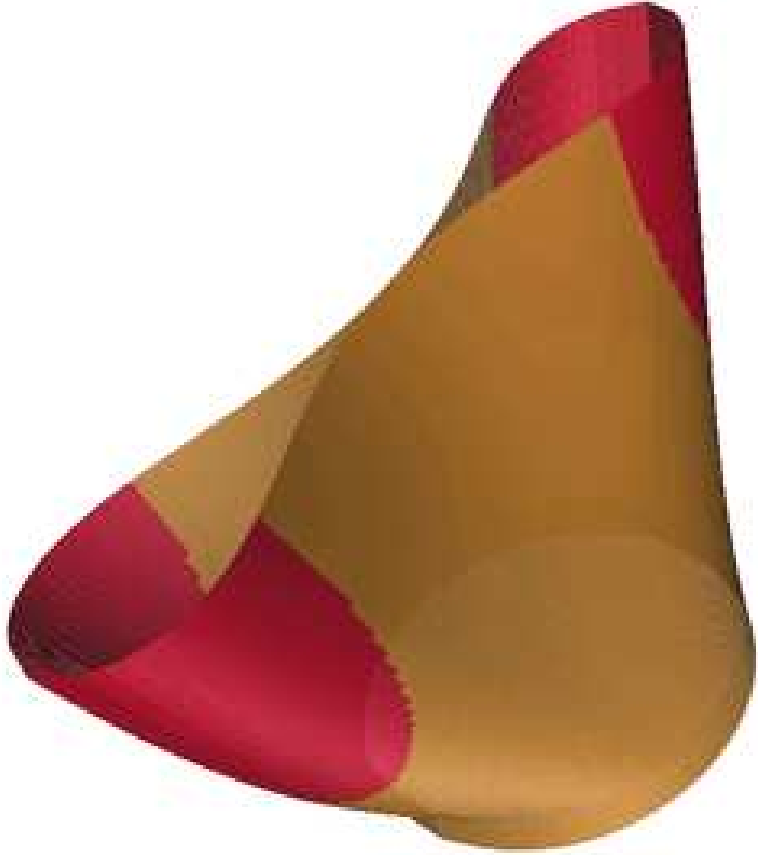}}
\caption{Embedding diagram for $\epsilon=10^{-5}$ to depict the shape of the
surface, the scales being arbitrary. The darkened region indicates
where the outward expansion has gone negative. The expansion first
reaches zero at the poles.}
\label{fig:eps-5_frontside}
\end{figure}

\begin{figure}
\centerline{\epsfxsize=6in\epsfbox{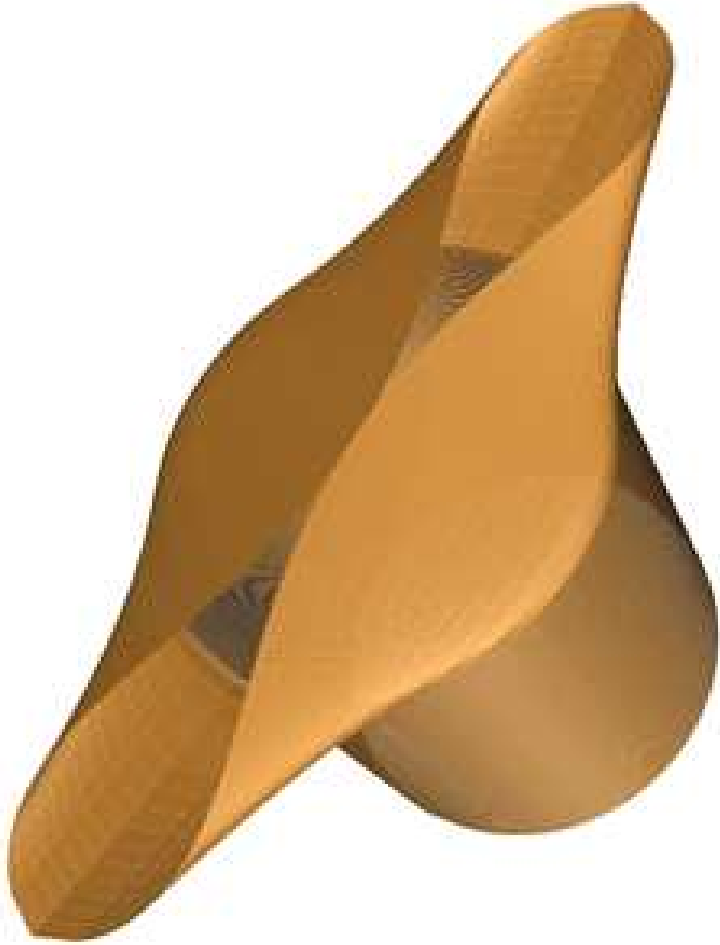}}
\caption{Embedding picture for $\epsilon=10^{-3}$. The radius reaches zero
first at the equator, at the center of the picture where the individual white
holes separate. The expansion is everywhere positive at that time.}
\label{fig:eps-3}
\end{figure}

Figures~\ref{fig:eps-5_frontside} and~\ref{fig:eps-3} are embedding
pictures of the horizon which reveal the angular behavior of the race.
(The construction of the embedding pictures is explained in
Ref.~\cite{asym}). The darkened portions of the pictures indicate where
the expansion has gone negative.  Figure~\ref{fig:eps-5_frontside}
shows that when the expansion reaches zero first it does so at the
pole; whereas the radius first reaches zero at the equator where the
horizon pinches off to form separate white holes.
Figure~\ref{fig:eps-3} shows that for $\epsilon = 10^{-3}$ the
pinch-off occurs before the outward expansion has gone to zero along
any ray.

More insight into the angular behavior is provided by surface plots as
functions of $(u,\theta)$ of the quantities $r$, $\Theta_{OUT}$, the
curvature scalar ${\cal R}$, the twist (as described by the normalized
component $\omega_{\hat \theta} = \omega_a \hat \theta^a$, ), and the
``plus'' component of the outgoing shear

\begin{equation}
        \sigma_{+, OUT} = \frac{1}{2} (\hat \theta^a \hat \theta^b
            -\hat \phi^a \hat \phi^b) \nabla_al_b
          = \frac{1}{4} \left[ \ln\left(\frac{K + \Re(J)}{K - \Re(J)}
                 \right)\right]_{,\lambda}   ,
\end{equation}
where $\hat
\theta^a$ and $\hat \phi^a$ are unit vectors in the $\theta$ and $\phi$
directions.
(The remaining components of the twist and shear vanish because of symmetry.)
Because of axial and
reflection symmetry, the range $0\le \theta \le \pi /2$ suffices to display the
full angular behavior.

\begin{figure}
\centerline{\epsfxsize=6in\epsfbox{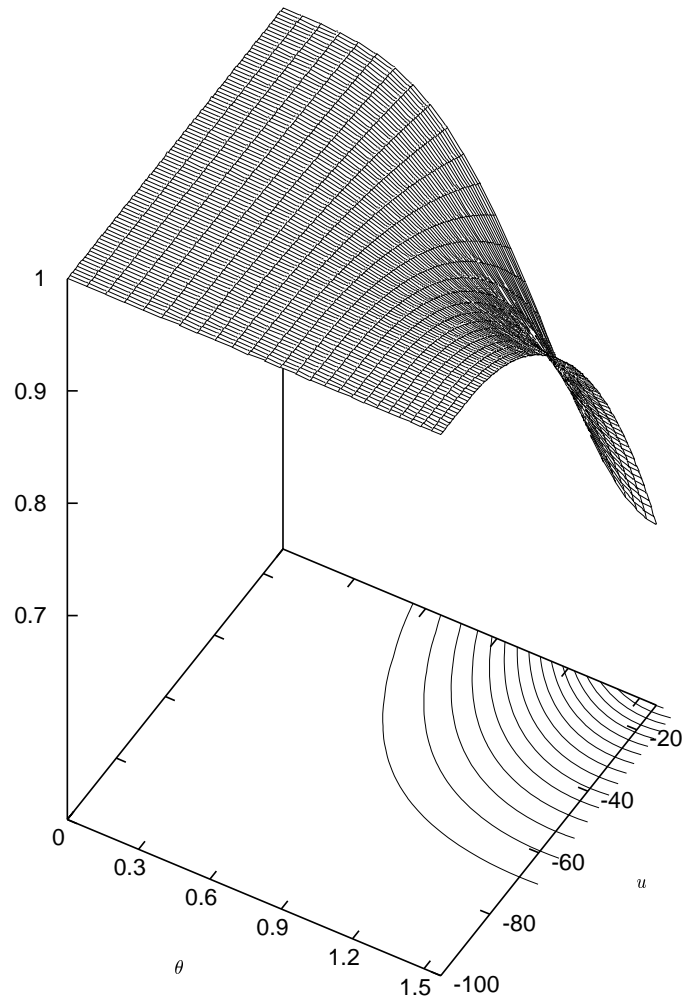}}
\caption{The radius decreases fastest at the equator,
in accord with the trouser-shaped horizon picture.}
\label{fig:rho5}
\end{figure}

\begin{figure}
\centerline{\epsfxsize=6in\epsfbox{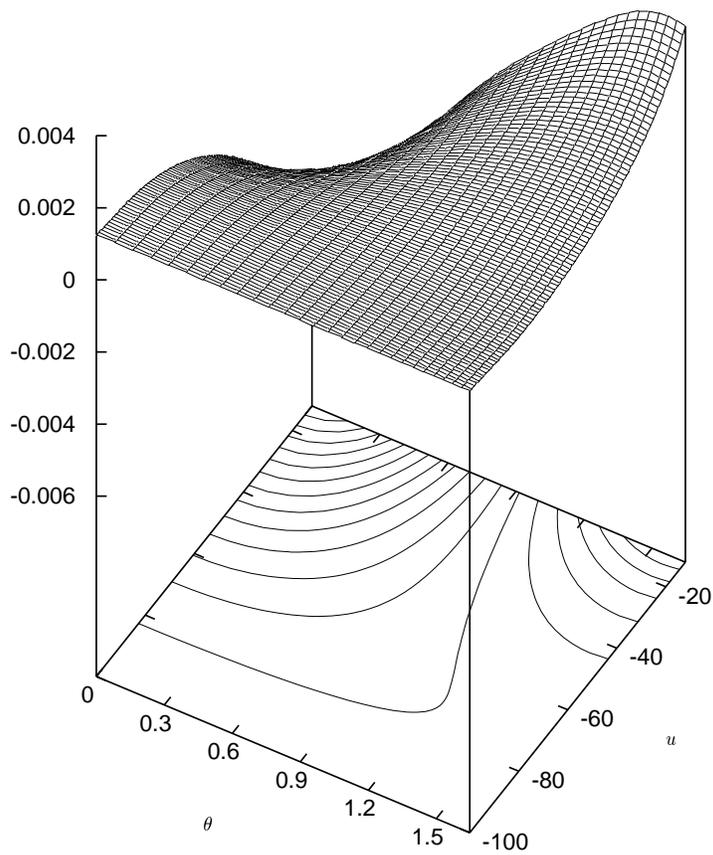}}
\caption{The minimum of the expansion occurs at a pole, hence
the expansion wins the race along a polar ray.}
\label{fig:exp5}
\end{figure}

\begin{figure}
\centerline{\epsfxsize=6in\epsfbox{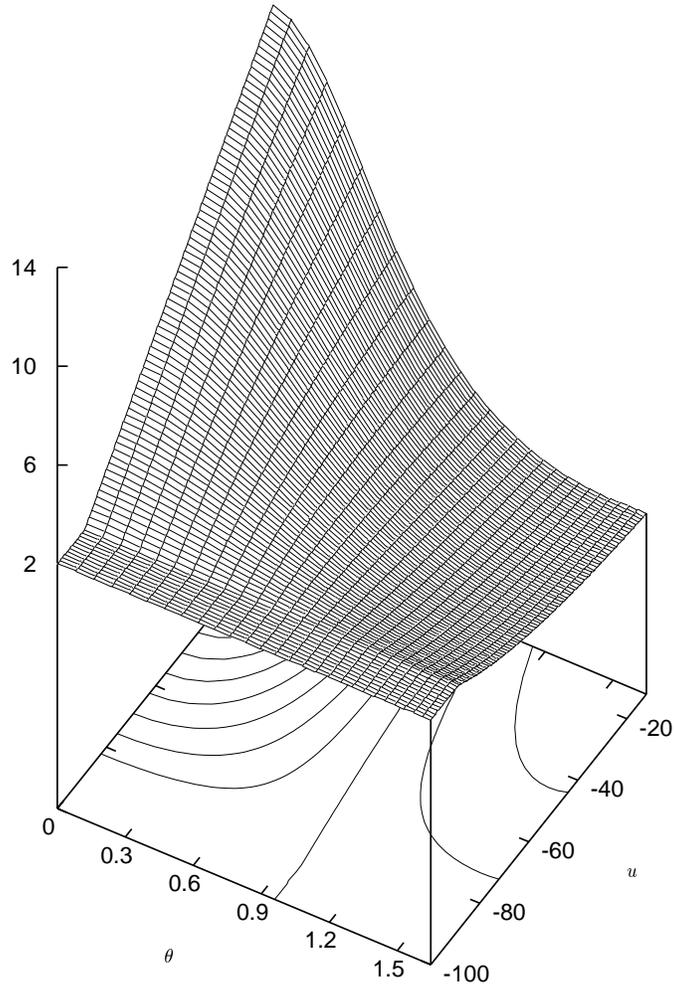}}
\caption{The curvature scalar ${\cal R}$ of the conformal horizon metric
$h_{AB}$ increases from its unit sphere value ${\cal R}=2$ in the region
near the poles and decreases in the region near the equator. }
\label{fig:ricci5}
\end{figure}

\begin{figure}
\centerline{\epsfxsize=6in\epsfbox{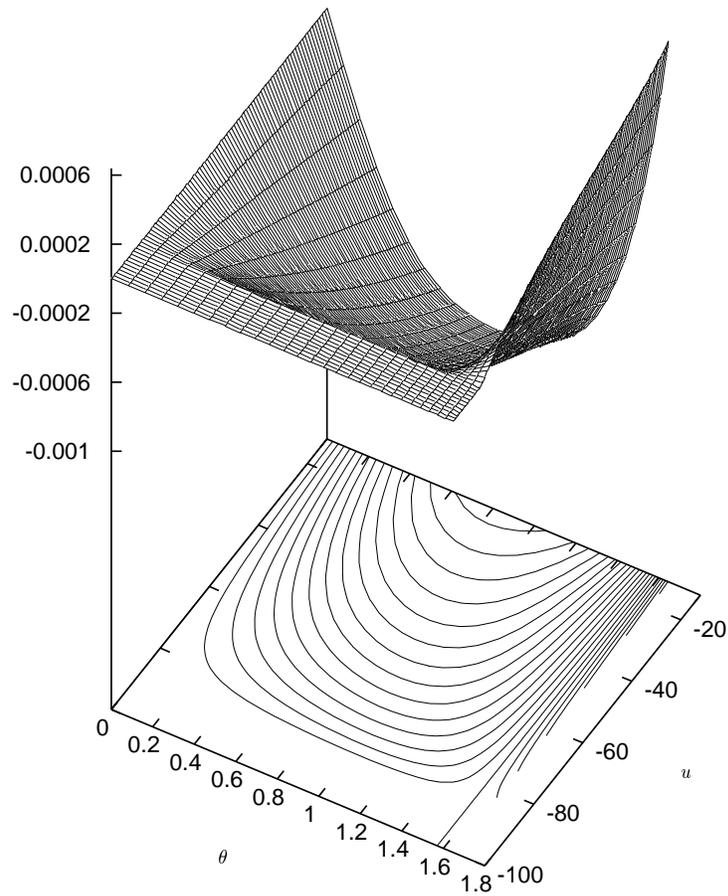}}
\caption{The behavior of the twist, a quantity which would vanish for
the Minkowski time slices of the flat space wave front, shown here for
$\epsilon=10^{-5}$.}
\label{fig:twist5}
\end{figure}

\begin{figure}
\centerline{\epsfxsize=6in\epsfbox{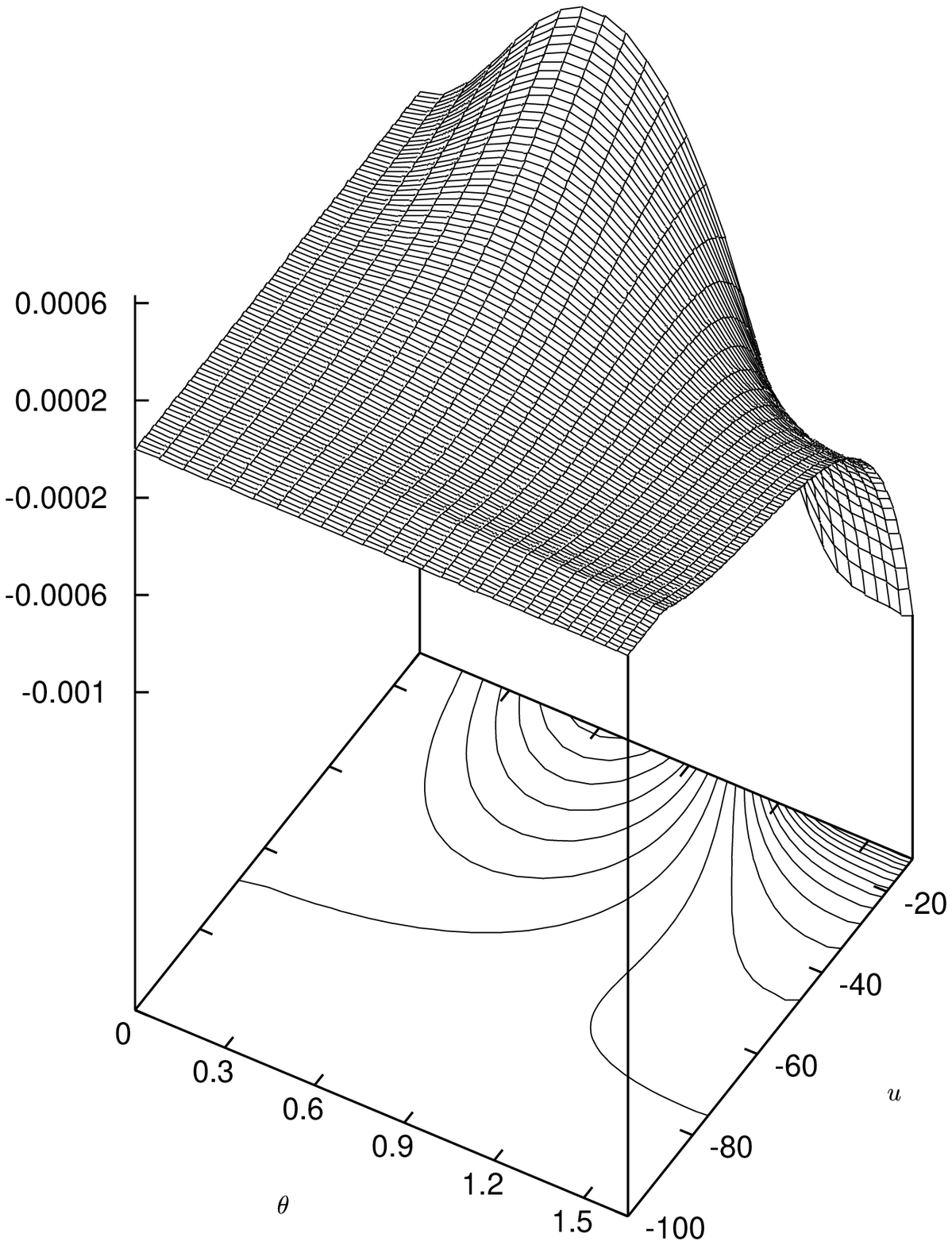}}
\caption{The shear $\sigma_{+,OUT}$ , as depicted in
Fig.~\ref{fig:shearp5}, is positive near the poles and negative near
the equator.}
\label{fig:shearp5}
\end{figure}

The first set of surface plots, Figs.~\ref{fig:rho5}-\ref{fig:shearp5}, are
for the mildly nonlinear case $\epsilon=10^{-5}$, corresponding to panel (c)
in Fig.~\ref{fig:exp-r-race} in which the expansion wins the race. The
evolution is traced from the starting time to the finish when the Bondi
boundary forms. Figure~\ref{fig:rho5} shows that the radius decreases fastest
at the equator, in accord with the trouser-shaped horizon picture.
Figure~\ref{fig:exp5} shows that the expansion wins the race along a polar
ray.  Figure~\ref{fig:ricci5} shows that the curvature scalar ${\cal R}$ of
the  conformal horizon metric $h_{AB}$ increases from its unit sphere value
${\cal R}=2$ in the region near the poles and decreases in the region near
the equator. This is what would be expected for the conformal geometry of the
Minkowski time slices $\hat t =const$ of of the collapsing prolate wave front
which seeds the model. However, it is important to bear in mind that the
quantities in Figs.~\ref{fig:rho5}-\ref{fig:shearp5} refer to the curved space
affine time slices $t=const$. This is emphasized in Fig.~\ref{fig:twist5},
which shows the behavior of the twist, a quantity which would vanish for the
Minkowski time slices of the flat space wave front. The shear
$\sigma_{+,OUT}$, as depicted in Fig.~\ref{fig:shearp5}, is  positive near the
poles and negative near the equator, again as would be  expected from a
prolate Minkowskian wave front (according to our above polarization
convention). Although the time dependence of all quantities in the model is
analytic, there is an apparent ``crease'' at $t\approx =-80$ in the surface
plots of the Ricci scalar, twist and shear which results from a rapid change
in the quantity  $\Lambda^{-1}=dt/d \hat t$ governing the relative Minkowski
and curved space affine times. For models with larger $\epsilon$, the angular
dependences are qualitatively similar but the time dependence is more
dramatic.

\section{Discussion} \label{sec:discussion}

Expressed now in terms of a black hole merger, this work has traced the horizon
structure of a head-on black hole collision from the close approximation to the
nonlinear regime. It has revealed dramatic time dependence in the intrinsic and
extrinsic curvature properties of the horizon in the extreme nonlinear regime.
The results suggest two classes of binary black hole space-times depending upon
whether the crotch in the standard trouser picture is protected, in the sense
that it lies inside a marginally anti-trapped surface on the horizon, or bare.
Only in the bare case is it possible that the black holes are formed by either
the collapse of matter or the implosion gravitational waves (see Fig.
\ref{fig:strategy}) originating in an initially nonsingular space-time.

The results pave the way for an application of the PITT code to calculate the
fully nonlinear wave forms emitted in the merger to ringdown phase. It remains
to be seen in future work whether the dramatic time dependence in the merger
stage is responsible for equally dramatic wave forms.

While the numerical results presented in this paper are for the axially
symmetric case, the codes are not restricted to any symmetry. It will be
interesting to see how the results for a head-on collision are modified in the
inspiral and merger of spinning black holes.

\acknowledgements

This work has been partially supported by NSF PHY 9510895 and NSF PHY
9800731 to the University of Pittsburgh. We have benefited from
conversations with our longtime collaborators Luis Lehner and Nigel T.
Bishop. R.G. thanks the Albert-Einstein-Institut for hospitality.
Computer time for this project was provided by the Pittsburgh
Supercomputing Center and by NPACI.

\appendix

\section{Numerical integration}
\label{app:numint}

\subsection{Integration of $\rho$}

As explained in Sec.~\ref{sec:confmod}, the conformal model supplies the area
coordinate $r$ as well as the null data $J$, but for completeness we describe
here the integration of the Raychaudhuri equation~(\ref{eq:ruu}) that
determines $r$ from the null data in a more general setting. We integrate this
equation in terms of the variable $\rho=r/r_M$, where $r_M$, defined in Eq.
(\ref{eq:rm}), satisfies $\dot r_M{}_{|\lambda=0}=0$ and $\ddot r_M =0$. As a
result, the evolution equation for $\rho$ is identical to that for $r$, which
in the spin-weighted form of Eq.~(\ref{eq:ruus}) becomes
\begin{equation}
      \ddot \rho =-\frac{\rho}{4} (\dot J \dot {\bar J}-\dot K^2) .
\end{equation}
The initial data consists of the values of $\rho$ and $\dot \rho$ on
the initial slice. Note that $\dot K = \Re(\dot J \bar J) / K$.
We put the equation in first-order form,
\begin{equation}
 \dot \rho = \Pi, \quad
 \dot \Pi = -4\, S \rho
\label{fosrho}
\end{equation}
with $S=(\dot J \dot {\bar J}-\dot K^2)/16$.

The advantage of the first-order form is that the pair of equations
(\ref{fosrho}) can be discretized to second-order accuracy using only
two time levels, in the same footing as the other horizon evolution
equations. The time integration stencil is the midpoint rule~\cite{recipes},
\begin{mathletters}
\begin{eqnarray} \label{rhostencil}
\frac{\rho^{n+1} - \rho^{n}}{\Delta u} &=& \frac{1}{2}
   (\Pi^{n+1} + \Pi^{n})
\\
\frac{\Pi^{n+1} - \Pi^{n}}{\Delta u} &=& - 2\, S^{n+\frac{1}{2}}
   (\rho^{n+1} + \rho^{n})
\end{eqnarray}
\end{mathletters}
which can be solved simultaneously for $\rho^{n+1}$ and $\Pi^{n+1}$
to give
\begin{mathletters}
\begin{eqnarray} \label{fosrhofde}
\rho^{n+1} &=& \frac{\rho^{n} (1 - S {\Delta u}^2) + \Pi^{n} {\Delta u}}
                    {1 + S {\Delta u}^2}
\\
\Pi^{n+1} &=& \frac{\Pi^{n} (1 - S {\Delta u}^2)
                        + 4\, S \rho^{n} {\Delta u}}
                      {1 + S {\Delta u}^2}.
\end{eqnarray}
\end{mathletters}

\subsection{Integration of $\omega$}

The time dependence of $\omega$ is determined by Eq.~(\ref{eq:omegadots}),
which we renormalize by factoring out $r_M^2$ from both sides. On the left
hand side, this is accomplished by using the identity
\begin{equation}
   (r^2 \omega)\dot{} = r_M^2 ( \rho^2 \omega ) \dot {},
\end{equation}
which holds on the horizon. Similarly, we re-express the first term on the 
right-hand side as
\begin{equation}
   r^2 \eth \left(\frac{\dot r}{r} \right) = r_M^2 \rho^2 \eth
  \left( { {\dot r_M \rho + r_M \dot \rho} \over {r_M \rho} } \right) =
  r_M^2 \rho^2 \left( \eth \left(\frac{\dot r_M}{r_M} \right)
  + \eth \left(\frac{\dot \rho}{\rho} \right) \right)
     =r_M^2 \rho^2 \eth \left(\frac{\dot \rho}{\rho} \right).
\end{equation}
Equation~(\ref{eq:omegadots}) then reduces to
\begin{mathletters}
\begin{eqnarray}
  (\rho^2 \omega)\dot{} &=&
  \left(\frac{1}{4} P_2 \rho + P_1 \right) \rho + P_0, \quad {\rm where}
\label{eq:dotomega} \\
   P_0 &=& - \dot\rho \eth\rho \\
   P_1 &=& \eth\dot\rho
         + (\dot J \bar J - \dot K K) \eth\rho
         + (J \dot K - K \dot J) \bar\eth\rho \\
   P_2 &=& \bar{\dot J} \eth J - 4 \dot K \eth K
         + 2 \dot K \bar\eth J + 3 \dot J \eth \bar J
         - 2 \dot J \bar\eth K + 2 J \bar\eth\dot K
         + 2 \bar J \eth\dot J - 2 K \bar\eth\dot J
         - 2 K \eth\dot K .
\end{eqnarray}
\end{mathletters}
We use the midpoint rule to integrate Eq.~(\ref{eq:dotomega}),
i.e. the left-hand side is evaluated as
\begin{equation}
  (\rho^2 \omega)\dot{} = \frac{1}{\Delta t} \left( (\rho^2 \omega)^{n+1}
                              - (\rho^2 \omega)^{n} \right)
\end{equation}
and $\rho$, $J$, etc. in $P_0$, $P_1$ and $P_2$ are evaluated at
$t^{n+1/2}$, {\it e.g.}
\begin{equation}
  J \equiv \frac{1}{2}(J^{n+1} + J^{n}), \quad
  \dot J \equiv \frac{1}{\Delta t}(J^{n+1} - J^{n}) .
\end{equation}
[Since $\omega$ does not enter in the right-hand side, this is a special case
of the second order Runge-Kutta scheme~\cite{recipes}, also used below in
Eqs.~(\ref{eq:rholu}) and (\ref{eq:jlrho}).]

\subsection{Integration of $\rho_{\lambda}$}

The time dependence of $r_{,\lambda}$, determined by Eq.~(\ref{eq:rlams}),
is re-expressed in terms of $\rho_{,\lambda}$ using the ansatz $r=r_M \rho$,
which yields, for $\lambda=0$,
\begin{equation}
   (r^2)_{,\lambda u} =
     8 M^2 \rho \rho_{,\lambda u}
   + (8 M^2 \rho_{,\lambda} - 2 u \rho ) \rho_{,u} - \rho^2 .
\end{equation}
Substitution into Eq.~(\ref{eq:rlams}) then gives
\begin{eqnarray}
   8 M^2 \rho \rho_{,\lambda u}
   + (8 M^2 \rho_{,\lambda} - 2 u \rho ) \rho_{,u} & = &
    \Re \left[
      \left( \bar\eth K - \eth \bar J \right)
      \left( \frac{\eth\rho}{\rho} + \omega \right)
              + K \left( \bar\eth\omega + \frac{\bar\eth\eth\rho}{\rho}
                        - \frac{\eth\rho\bar\eth\rho}{\rho^2}
                  \right) \right. \nonumber \\
              &-& \left. \bar J \left( \eth\omega + \frac{\eth^2\rho}{\rho}
                        - \frac{(\eth\rho)^2}{\rho^2}
                       \right)
       - \bar J \omega^2
      +   K \omega \bar\omega - \frac{1}{2}{\cal R} + \rho^2 \right],
\label{eq:rholu}
\end{eqnarray}
which we integrate using a second order Runge-Kutta scheme.

\subsection{Integration of $J_{\lambda}$}

After setting $r=r_M \rho$, the evolution equation~(\ref{eq:jlamdots})
for $J_{,\lambda}$ becomes
\begin{eqnarray}
   8 M^2 \rho^2 J_{,\lambda u} - u \rho^2 \dot J
   + \left(1 + K^2 \right) \eth \omega
 = - 8 M^2 \rho \left( \dot\rho J_{,\lambda} + \rho_{,\lambda} \dot J \right)
   + 4 M^2 J \left( \dot{\bar J} J_{,\lambda}
                       + \dot J \bar J_{,\lambda}
                       - 2\, \dot K K_{,\lambda}
                \right) \nonumber \\
+ \left(1 + K^2 \right)
   \left( \omega^2 - 2\, \omega \frac{\eth \rho}{\rho} \right)
       - \omega \left(J \bar\eth K - K \bar\eth J \right)
       - \bar\omega \left(K \eth J - J \eth K \right)
\nonumber \\
       + J \left(J \bar\eth \bar\omega
                   - K \left( \eth\bar\omega + \bar\eth \omega \right)
                   + J \bar\omega^2
                   - 2 K \omega \bar\omega
                   + 2\, \frac{K}{\rho} \left(  \omega \bar\eth\rho
                                              + \bar\omega \eth\rho
                                        \right)
                   - 2\, J \bar\omega \frac{\bar\eth\rho}{\rho}
             \right)
\label{eq:jlrho}
\end{eqnarray}
which we again integrate using a second order Runge-Kutta scheme.

\section{The case of axial symmetry}
\label{app:axial_symmetry}

In axial symmetry the Sachs metric Eq.~(\ref{eq:amet}) can be written as
\begin{equation}
       ds^2  = -(W - U^2 \gamma r^{-2})du^2 - 2 du d\lambda -2 U
          du\, dx + r^2\left( \frac{dx^2}{\gamma} + \gamma \,
           d\varphi^2\right).
\label{eq:aximet}
\end{equation}
Here $W$, $U$ and $r$ are functions of
$(u,\lambda,x=\cos{\theta})$. With the dyad choice associated with Eq.
(\ref{eq:headonj}) the conformal horizon model gives
\begin{equation}
     \gamma =(1-x^2)\Gamma,\quad \Gamma =
      \left(\frac  {\hat t - r_\phi} {\hat t
                -r_\theta} \right),\quad J = \frac{1}{2}\left(1/\Gamma
                    - \Gamma\right),
\label{eq:gamma_and_J}
\end{equation}
with $\Gamma$ a well-behaved function on
the sphere. For the prolate spheroidal model considered here,  $\Gamma$
vanishes at the points $\hat t =r_\phi$ where the horizon pinches off and
$\Gamma =O(r^2)$.
Thus, in the prolate case, it is useful for
numerical purposes to use $\gamma$, as opposed to $\gamma^{-1}$, as the
variable to represent the metric on the two-sphere of constant $u$ and
$\lambda$. The twist $\omega$ is related to the quantity $U$ by
\begin{equation}
         \omega = -\frac{1}{2} \sqrt{1 - x^2} \, U_{,\lambda} .
\label{eq:capU}
\end{equation}

The Einstein equations yield the following system of PDE's for propagating
the metric variables along the horizon:
\begin{eqnarray}
    \partial_u (r^2 U_{,\lambda}) &=&
        2\,r_{,u}\,r_{,x} - 2\,r\,r_{,xu} -
           \frac{2\,r\,r_{,x}\,\gamma_{,u}}{\gamma}
           - \frac{r^2\,\gamma_{,xu}}{\gamma}
\label{eq:dot_r2Ul}\\
     \partial_u \partial_\lambda r^2 &=&  -\frac{(r_{,x})^2\,\gamma}{r^2} +
           \frac{r_{,xx}\,\gamma}{r} + \frac{(U_{,\lambda})^2 \,\gamma}{4} -
       \frac{U_{,\lambda  x}\,\gamma}{2} + \frac{r_{,x}\,\gamma_{,x}}{r} -
       \frac{U_{,\lambda}\,\gamma_{,x}}{2} + \frac{\gamma_{,xx}}{2}
\label{eq:dot_r2l}\\
         \partial_u \partial_\lambda \, \gamma &=&
      \frac{r_{,x}\,U_{,\lambda}}{r^3}
       + \frac{(U_{,\lambda})^2 - 2 U_{,\lambda x} }{4 r^2}
       +\frac{\gamma_{,u}\,\gamma_{,\lambda}}{\gamma}
     -\frac{r_{,\lambda}\,\gamma_{,u} + r_{,u}\,\gamma_{,\lambda}}{r} .
\label{eq:dot_gammal}
\end{eqnarray}

In the Schwarzschild case, $r=r_M=2M -(\lambda u/4M)$, $U=0$ and $\gamma =
(1-x^2)$. Thus, if the geometry is near Schwarzschild, all terms on the
right-hand side are small except $\gamma_{xx}=-2$.
In order to make Eq.~(\ref{eq:dot_r2l}) numerically well behaved we subtract
this term by introducing the auxiliary variable $\Delta=\partial_\lambda (r^2
- r_M^2)$. Then $\Delta$ satisfies
\begin{equation}
      \partial_u \Delta =  -\frac{(r_{,x})^2\,\gamma}{r^2}
      + \frac{r_{,xx}\,\gamma}{r}
      + \frac{(U_{,\lambda})^2 \,\gamma}{4}
      - \frac{U_{,\lambda  x}\,\gamma}{2}
       + \frac{r_{,x}\,\gamma_{,x}}{r}
     - \frac{U_{,\lambda}\,\gamma_{,x}}{2}
      + \left(\frac{\gamma_{,xx}}{2} + 1\right) .
\label{eq:dot_Delta}
\end{equation}
The quantity $r_{,\lambda}$ is reconstructed as
\begin{equation}
r_{,\lambda} = \frac{\Delta + (r_M^2)_{,\lambda}}{2 r}
             = \frac{\Delta -                  u}{2 r}.
\end{equation}

Special care is needed in order to write the right-hand sides of
Eqs.~(\ref{eq:dot_r2Ul}) and (\ref{eq:dot_gammal}) in manifestly
regular form (for $r\neq 0$) at the axis of symmetry ($x=-1, 1$)
where $\gamma= (1-x^2)\,\Gamma$ vanishes.
We thus express $\gamma_{,u}/\gamma$ as $\Gamma_{,u}/\Gamma$ and
$\gamma_{,x u}/\gamma$ as
\begin{equation}
      \frac{\gamma_{,x u}}{\gamma}
     =  \frac{\Gamma_{,ux}}{\Gamma}
       - \frac{2x \Gamma_{,u} }{(1-x^2) \Gamma }  .
\end{equation}
Axial symmetry implies that $\Gamma_{,u}$ vanishes at the poles so that the
right-hand side of Eq.~(\ref{eq:dot_r2Ul}) is regular, as can be seen by
differentiating the conformal data in Eq.~(\ref{eq:gamma_and_J}) to obtain
\begin{equation}
       \Gamma_{,u} =- \left( \frac{\partial \hat t}{\partial u} \right)
         \frac{(r_\phi -r_\theta)}{ (\hat t -r_\theta)^2}
\end{equation}
and noting that $r_\phi =r_\theta$ on the symmetry axis. Thus Eq.
(\ref{eq:dot_r2Ul}) is rendered explicitly regular on the symmetry axis. At
the pinch-off points at $r=0$, Eqs.~(\ref{eq:dot_r2Ul}) -
(\ref{eq:dot_gammal}) are singular but the numerical performance near these
points can be enhanced by noting that  $\gamma/r^2$ remains regular, by virtue
of Eqs.~(\ref{eq:def_rhat}) and (\ref{eq:gamma_and_J}).

The data at ${\cal S}_-$ are initialized by the same prescription
as for the general case discussed in Sec.~\ref{sec:sminus}, which implies
\begin{equation}
    \partial_\lambda U_- = J_{-,x} - \frac{2 x}{1-x^2} J_- \, ,
\end{equation}
in accord with Eq.~(\ref{eq:omegaminus})
 \begin{equation}
           \Delta_- = u_- \,\left(1 - \frac{r_-^2}{4M^2}\right) ,
\end{equation}
in accord with Eq.~(\ref{eq:rhominus}) and
\begin{equation}
       \partial_\lambda \gamma_- = 0  .
\end{equation}
in accord with Eq.~(\ref{eq:jminus}).

We integrate Eqs.~(\ref{eq:dot_r2Ul}), (\ref{eq:dot_gammal}) and
(\ref{eq:dot_Delta}) numerically to second order accuracy, using the same
midpoint rule as for the general case in Appendix~\ref{app:numint}.

\section{Convergence tests}

We have verified second order convergence of the hierarchy of equations
which provide horizon data for the 3D code.

We have also checked second order convergence of the axially symmetric
code, and we have used it to confirm the behavior, described in
Sec.~\ref{sec:results}, of the sudden nonlinear plunge of the horizon
radius to zero. Our aim is to use the axially symmetric code to
generate an independent numerical solution against which to check the
3D code.

Due to its lower computational requirements, the axisymmetric
calculation is significantly more efficient than the 3D one.  This will
be particularly useful in the approach to the caustic, where the
highest resolution is needed, and where the axisymmetric code will
allow us to make more detailed studies of the behavior of the horizon.



\newpage

\begin{thebibliography}{10}

\bibitem{ndata}
Luis Lehner, Nigel T. Bishop, Roberto G\'omez, Bela Szil\'agyi, and
Jeffrey Winicour, Phys. Rev. D {\bf 60}, 044005 (1999).

\bibitem{asym}
Sascha Husa and Jeffrey Winicour, Phys. Rev. D {\bf 60}, 084019 (1999).

\bibitem{pp}
Richard H. Price and Jorge Pullin, Phys. Rev. Lett. {\bf 72}, 3297
(1994).

\bibitem{high}
Nigel T. Bishop, Roberto G\'omez, Luis Lehner, Manoj Maharaj, and
Jeffrey Winicour, Phys. Rev. D {\bf 56}, 6298 (1997).

\bibitem{wobb}
R. G\'{o}mez, L. Lehner, R.~L. Marsa, and J. Winicour, Phys. Rev. D
{\bf 57}, 4778 (1997).

\bibitem{kyoto}
J. Winicour, Prog. Theor. Phys. Suppl. {\bf 136}, 57 (1999).

\bibitem{close1}
Manuela Campanelli, Roberto G\'omez, Sascha Husa, Jeffrey Winicour,
and Yosef Zlochower, Phys. Rev. D (to be published), gr-qc/0012107.

\bibitem{bondi}
H. Bondi, M. van~der Burg, and A. Metzner, Proc. R. Soc. London, {\bf
A269}, 21 (1962).

\bibitem{sachs}
R. Sachs, Proc. R. Soc. London, {\bf A270}, 103 (1962).

\bibitem{sachsdn}
R. Sachs, J. Math. Phys. {\bf 3}, 908 (1962).

\bibitem{ih5083}
Abhay Ashtekar, Stephen Fairhurst, and Badri Krishnan, Phys. Rev. D
{\bf 62}, 104025 (2000).

\bibitem{haywsn}
S.~A. Hayward, Class. Quantum Grav. {\bf 10}, 773 (1993).

\bibitem{haywdn}
S.~A. Hayward, Class. Quantum Grav. {\bf 10}, 779 (1993).

\bibitem{newt}
J. Winicour, J. Math. Phys. {\bf 24}, 1193 (1983).

\bibitem{nullinf}
J. Winicour, J. Math. Phys. {\bf 25}, 2506 (1984).

\bibitem{competh}
R. G\'{o}mez, L. Lehner, P. Papadopoulos, and J. Winicour, Class. Quantum
Grav. {\bf 14}, 977 (1997).

\bibitem{penrin}
R. Penrose and W. Rindler, {\em Spinors and Space-Time} (Cambridge
University Press, Cambridge, England, 1984), Vol.~1.

\bibitem{sachseq}
R.~K. Sachs, Proc. R. Soc. London {\bf A264}, 309 (1961).

\bibitem{israel}
W. Israel, Phys. Rev. {\bf 143}, 1016 (1966).

\bibitem{bartnik1}
R. Bartnik and A.~H. Norton, in {\em Computational Techniques and
Applications: CTAC97}, edited by B.~J. Noye, M.~D. Teubner, and A.~W. Gill
(World Scientific, Singapore, 1998), p.\ 91.

\bibitem{bartnik2}
R. Bartnik and A.~H. Norton, SIAM J. Sci. Comput. (USA) {\bf 22}, 917 (2000).

\bibitem{pensing}
R. Penrose, Phys. Rev. Lett. {\bf 14}, 57 (1965).

\bibitem{recipes}
W. Press, B. Flannery, S. Teukolsky, and W. Vetterling, {\em Numerical
Recipes} (Cambridge University Press, New York, 1986), Chap.~6.6, p.\ 182.

\end{thebibliography}
\end{document}